\documentclass[aps,prb,twocolumn,floatfix,superscriptaddress,showpacs]{revtex4-1}
\usepackage{amsmath}
\usepackage{color}
\usepackage{ulem}
\usepackage{natbib}
\usepackage{enumerate}
\usepackage[dvips]{graphicx}

\begin{document}
\newcommand{\ra}{$t_h/t_l$}
\newcommand{\tup}{$t_l$}
\newcommand{\tdw}{$t_h$}
\newcommand{\up}{$l$}
\newcommand{\dw}{$h$}
\newcommand{\tn}{$T_o$}
\newcommand{\docc}{$\langle n_h n_l\rangle$}
\newcommand{\dav}{$\langle d\rangle$}
\newcommand{\cv}{$C_{\rm V}$}
\newcommand{\oop}{$m_o$}
\newcommand{\cdwop}{$m_c$}
\renewcommand{\theparagraph}{\roman{paragraph}}

\title{Phase diagram of the Asymmetric Hubbard Model and an  entropic chromatographic method for cooling cold fermions in optical lattices}
\author{E. A. Winograd}
\affiliation{Laboratoire de Physique des Solides, CNRS-UMR8502, Universit\'e de Paris-Sud,
Orsay 91405, France.}
\author{R. Chitra}
\affiliation{Laboratoire de Physique Theorique de la Mati\`ere 
Condense\'e, UMR 7600, Universit\'e de Pierre et Marie Curie, Jussieu, Paris-75005, France.}
\affiliation {Theoretische Physik, ETH Zurich, 8093 Zurich, Switzerland.}
\author{M.J. Rozenberg}
\affiliation{Laboratoire de Physique des Solides, CNRS-UMR8502, Universit\'e de Paris-Sud,
Orsay 91405, France.}

\date{\today}
\pacs{67.85.Lm, 71.30.+h, 71.10.Fd, 75.10.Jm}

\begin{abstract}
We study  the  phase diagram of the asymmetric Hubbard model (AHM),  which is characterized by different values of the hopping for the two spin projections of  a fermion or equivalently, two different orbitals. This model  is expected to provide a good description of
  a mass-imbalanced  cold fermionic mixture in a 3D optical lattice. We use the dynamical mean field theory to study  various
  physical properties of this system.  In particular, we show how orbital-selective physics, 
observed in multi-orbital  strongly correlated electron systems, can be realized in such a simple model.    We find that  the density distribution  is a good probe of this   orbital selective crossover from a Fermi liquid to a non-Fermi liquid state. 
 Below an ordering temperature $T_o$, which is a function of both the interaction and  hopping asymmetry, the system  exhibits  staggered long range orbital order. Apart from the special case of the  symmetric limit, i.e., Hubbard model, where there is no hopping asymmetry,
this orbital order is accompanied by  a true charge density wave order for all values of the hopping asymmetry.
We calculate the order parameters  and various physical quantities including the thermodynamics in both the ordered and
disordered phases. We find that the formation of the charge density wave is signaled by an abrupt increase in the sublattice double occupancies. 
Finally, we  propose a new method, entropic chromatography,  for cooling  fermionic atoms in optical lattices, by exploiting the properties of the AHM.  To establish this cooling strategy on a firmer basis, we also discuss the variations in temperature induced by the adiabatic  
 tuning of  interactions and hopping parameters.
\end{abstract}

\maketitle

\section{Introduction}
Strongly correlated  electronic systems  are known to exhibit a wide variety of  interesting and novel phases  ranging from Fermi liquids, heavy fermions, 
Mott insulators, superconductors  to other states with complex long range orders. 
The field of strongly correlated fermions in condensed matter  shares many common research goals with that of  cold atoms loaded in optical lattices.
Yet these two fields emerge from different scientific communities and often look at similar problems with different tools
and from different physical perspectives. In this context, the present theoretical study focuses on a model hamiltonian 
that generalizes central models of strongly correlated electron systems, which, nevertheless, might only be experimentally realized
with  cold atoms in an  optical lattice. The present  work is a contribution towards building bridges between these two
scientific communities. 

The paradigmatic  model of correlated fermions is 
the Hubbard model\cite{hubbard-orig} (HM), which describes itinerant  electrons  on a lattice with hopping $t$ and subjected to  an onsite  repulsion $U$
stemming from  screened Coulomb interactions between two electrons.
This model,  though the focal point of interest in condensed matter physics for decades, however 
remains unsolved in two and three  spatial dimensions.  In
 the limit of infinite dimensions, or infinite lattice connectivity,  the model remains non-trivial but is exactly solvable using the dynamical mean field theory (DMFT)\cite{bible}. 
DMFT shows that  in the  absence of any frustration, the ground state of the system at half-filling is an 
antiferromagnetic insulator for all values of $U$ with an associated N\'eel temperature. 
In cases where due to inherent frustration the system is in a paramagnetic  state,  one finds a correlated metal
described by a Fermi liquid with heavy quasiparticles  at  low and moderate $U$. Upon increasing the interaction
strength, always at half-filling, the heavy metal breaks down and a Mott insulating  
state is realized with a coexistence region \cite{bible}. 
Such a  Mott transition was indeed observed in the classic  three dimensional Mott-Hubbard system V$_2$O$_3$ \cite{bible}.

Another important model Hamiltonian of strongly correlated systems is the Falicov-Kimball model\cite{fk-orig} (FKM),
which is a simplified variant of the HM where one of the spin species has zero hopping, hence, its motion
is frozen.   This model explicitly breaks the spin
SU(2) symmetry and
DMFT studies \cite{fk} showed that it has an antiferromagnetically  ordered insulating phase at half-filling, and regions of phase 
separation at low values of the onsite Coulomb interaction for finite doping.  
Similar to the HM, a paramagnetic Mott metal -insulator transition (without any coexistence) as a function of the interaction strength was also found in the FKM at
half filling.   A fundamental difference between the two models is that while the
metallic state in the HM was described by a Fermi liquid, the metallic state in the FKM is   a non-Fermi liquid with no quasiparticles.  This exotic state  stems  from the broken translational
symmetry due to the random distributions of the immobile electrons on the lattice. \cite{bible,sietal,fk}.

The natural connection between these two models is the Asymmetric Hubbard model (AHM), where  each spin species
has a different hopping amplitude.  This model,  originally considered purely academic in  solid state systems,  is of
particular relevance  to cold atoms on optical lattices. In fact, asymmetric Hubbard models  are realizable in optical lattices, using either two hyperfine states of a 
spin-$1/2$ atom or two different spinless fermionic atoms (e.g., $^6\mathrm{Li}$ and $^{40}\mathrm{K}$), as described in Refs. \onlinecite{blochprl91,japanese,rmpcoldatoms,giamarchi,sarma,dao,lda}. 
Each of the hopping amplitudes $t_\sigma$ can be independently controlled by changing the intensity of the lasers and the Coulomb interaction strength $U$ 
is related to the scattering length and can be well controlled by  Fano-Feshbach resonance\cite{PhysRevLett.100.053201,RevModPhys.feshbach}. 
Following Refs. \onlinecite{salomon,Capone}, the hopping asymmetry $r$ in a mixture of $^6\mathrm{Li}$/$^{40}\mathrm{K}$ can be tuned from values close to the FKM (r=0) to
that of the HM (r=1).

Another important reason to study the AHM, is the fact that it can be thought of as a minimal model of multi-orbital systems. Earlier multi-orbital models in (bosonic) cold-atom systems have been considered in Refs. \onlinecite{ref1,ref2,ref3}. In general, in correlated materials, often when there is more than one band close to the Fermi energy, new physical phenomena arise. The existence of many interacting bands close to  the Fermi energy, may generate exotic
phases, produced by the difference in the population of each band, the inter and intra band Coulomb interaction, as well as the Hund's coupling. 
Examples of these materials, 
which are subjects of current research include high-T$_\textrm{c}$ iron-based superconductors, where 5 bands cross the Fermi energy. From this perspective, the AHM, which can be viewed  as a model for spinless fermions
with two different orbitals, can be considered as an attempt to approach the physics  of 
correlated multi-orbital systems, in cold-atoms.
For this reason, in this paper we adopt the notation of  two 
different orbitals $h$ and $l$, instead of two spin-projections $\uparrow$ and 
$\downarrow$, as we will explain later.

Most of the recent theoretical work on the AHM in the cold-atoms context, has
concentrated on establishing the phase diagram in the one dimensional case, for both, attractive and repulsive interactions.
A rich variety of groundstates have been found, including Mott insulator, charge density wave and superconductivity, in
the case of mixtures of equal density \cite{giamarchi}, and also FFLO states in the case of population imbalance \cite{sarma}.
Beyond the one dimensional case, the AHM  with attractive interactions  on  a cubic optical lattice, has been studied using DMFT  \cite{dao}.  A related, but more general model,
was also analyzed in Ref. \onlinecite{brydon}  using the slave-boson technique.
The phase diagram of the repulsive AHM without any long-range order was studied using DMFT in Refs.\onlinecite{emilio_prb, Capone}.
Both studies found  a Mott transition with a region of two coexistent solutions, at low enough temperatures for all non-zero values of the hopping asymmetry. 
In particular, in Ref.\onlinecite{emilio_prb},  we showed that an interesting temperature driven orbital-selective crossover takes place in the AHM. At  temperatures  below 
a coherence temperature $T_{coh}$, the system is a heavy Fermi-liquid, qualitatively analogous to the Fermi liquid state of the symmetric Hubbard model. Above  $T_{coh}$,  
instead of the standard incoherent metal seen in the HM case, an orbital-selective crossover occurs, wherein one fermionic species effectively localizes, and the metallic 
state maps onto the non-Fermi liquid of the Falicov-Kimball model.  This is a validation of the fact that the AHM can be
considered as a minimal model for the study of orbital physics.  
Various  observables such as the double occupation, the specific heat and entropy  were also computed.
 As stated earlier, a phase diagram without any long-range order is appropriate  provided  an inherent 
 frustration prevents the formation of any kind of orbital ordering. 
However, for the unfrustrated case, which corresponds to the bare model hamiltonian without
additional assumptions, one needs to explore ordered  solutions to the mean field equations.  Here, for simplicity, we restrict ourselves to the case of half-filling,
where only N\'eel-type (ie checkerboard) order is realized. However, away from half-filling, incommensurate order may
also occur \cite{fk,camjayi}.  
To explore the long-range ordered solutions and to understand the way how these are connected to the
previously studied finite temperature crossover regimes, including the orbital selective state\cite{emilio_prb}, is the main goal of the present work. 
In this paper,  we  study the orbital-ordered phases as a function of hopping asymmetry, interaction strength and temperature, 
which then permits us to obtain a complete 
phase diagram of the half-filled AHM. 
Since  fermionic cold atom systems have not yet attained the low temperatures which are requisite for observing any
kind of order, we also propose a novel cooling scheme which is based on the hopping asymmetry of the two orbitals.

The present problem is at the interface between two different fields, thus, we summarize all the technical details in a self-contained section 
\ref{subsec:mfeq} that can be skipped if desired.  For further information on the DMFT-technique we suggest Ref. \onlinecite{bible}. 
Qualitative features of the long-range order phase are described in \ref{subsec:longrangequal}.
In section
\ref{subsec:pmphase}, we briefly review the results of the disordered phase of the AHM found in Ref. \onlinecite{emilio_prb}, and discuss their consequences
for cold atoms experiments. 
In section \ref{subsec:orderphase}, we discuss the ordered phase of the AHM and present results for the critical temperature $T_o$ , the different order parameters and  
the thermodynamical quantities 
of the model.
In section \ref{subsec:cooling}, we propose a strategy for cooling down a mixture of cold fermionic atoms, that exploits
the mass imbalance and may have practical applications.
In section \ref{subsec:coolingbytuning}, we analyze the effects on the 
temperature of the condensate of tuning the values of hopping and interaction of the atomic particles. 
Finally in section \ref{sec:conclusions}, we summarize our conclusions.
The expressions used to compute the physical observables can be found in the appendix.

\section{Model and Method}
\label{sec:method}
\subsection{Mean-field equations}
\label{subsec:mfeq}

The asymmetric Hubbard model Hamiltonian reads,
 \begin{equation}
\label{eq:ahm}
H=-\sum_{<i,j>;\sigma=l,h} t_\sigma c_{i\sigma}^\dag c_{j\sigma}^{\phantom{\dag}}+U\sum_i (n_{il}-1/2)(n_{ih}-1/2)
\end{equation}
\noindent
where $c^\dag_{i\sigma},c_{i\sigma}$ are the fermion creation and annihilation 
operators  at site $i$ and $\sigma=l,h$ labels  the two orbitals, light and heavy.
$t_\sigma$ and $U$ denote the hopping and the repulsive interaction strengths and the asymmetry parameter  
is defined as $r$=\ra. The limits $r$=0 and $r$=1 correspond to the FKM and HM, respectively.
 We restrict our study to the particle-hole symmetric  half-filled case, with  number of particles per site
$\langle n_{l} \rangle = \langle n_{h} \rangle = 1/2$. 
To describe a realization of the AHM  in  
cold atom systems in an optical lattice, one would need to add a confining potential to 
the hamiltonian \eqref{eq:ahm}. For simplicity, in this initial study, 
we do not include any confining potential in our hamiltonian. 
Nevertheless, based on previous results obtained for
the Hubbard model within the LDA or R-DMFT treatment including the trapping potential \cite{dao,prl105_germans}, we expect our results to be reliable for the bulk of the optical lattice.
The confinig potential typically affects the sites in a thin outer shell of the lattice, with a thickness of 
only a few sites\cite{prl105_germans}.

We now briefly discuss the DMFT method used to study the AHM.
This method, which is exact in the limit of infinite lattice coordination,  relies on a mapping of the present problem 
 onto a quantum impurity  problem  subject to self-consistency conditions. 
Here we consider the AHM on a  Bethe lattice 
 characterized by the  semi-circular density of states,
\begin{equation}
\rho_\sigma(\epsilon)=\frac{2}{\pi D_\sigma}\sqrt{1-(\epsilon/D_\sigma)^2}
\end{equation}
 with $D_\sigma=2t_\sigma$. We set  $D_l\equiv D=1$ as the unit of energy in what follows. The advantage of the
 Bethe lattice is that it  provides numerical simplifications  and also is expected to yield
accurate estimates for a 3D simple-cubic lattice\cite{bible}, adopting $D_\sigma=6t_\sigma$.

We solve the DMFT equations using two standard numerical methods: the Hirsch-Fye 
quantum Monte Carlo algorithm (HF-QMC), which is exact in the statistical sense; and 
the exact diagonalization method (ED), which is in principle exact and whose
discretization errors can be systematic reduced. Their implementation is described in detail
in Ref. \onlinecite{bible}. The use of these two different methods further provides a crosscheck 
of our numerical results.

To obtain a complete phase diagram of the  AHM, we need to consider both ordered 
and disordered  solutions to 
the dynamical mean field equations.  We discuss the two cases below.

\subsubsection{Disordered solution}

 In the disordered  phase, the system can be mapped onto a quantum
impurity problem  described by the  imaginary time effective action:
\begin{widetext}
\begin{equation}
\label{eq:impproblem}
\textrm{S}_\textrm{eff}=-\int_0^\beta d\tau \int_0^\beta d\tau' 
\sum_{\sigma=l,h} c^\dag_\sigma(\tau)
\mathcal{G}_{0\sigma}^{-1}(\tau-\tau')c_\sigma(\tau') + U\int_0^\beta d\tau
(n_l(\tau)-1/2)(n_h(\tau)-1/2)
\end{equation}		
\end{widetext}
The retarded Green's function $\mathcal{G}_{0\sigma}$ is related to the full local Green's function
$G_\sigma$ through the following self consistency conditions,
\begin{equation}
\label{eq:disordersc}
\mathcal{G}_{0\sigma}^{-1}(i\omega_n)=i\omega_n+\mu_\sigma -t_\sigma^2G_\sigma(i\omega_n)
\end{equation}
$\omega_n$ are the Matsubara frequencies and $\mu_\sigma$ are the chemical potentials for the two species.  
At the self consistent point, the impurity and the lattice local Green's functions
coincide. Therefore, the Eqs. \eqref{eq:impproblem} and \eqref{eq:disordersc}
can be simply solved by direct substitution. 
Note that unlike the HM case, here the disordered phase is characterized by two self-consistent
equations because the AHM explicitly breaks {\it orbital} rotational invariance  for all $r \ne 1$.

\subsubsection{Checkerboard-like ordered solution}
\label{subsec:longrange}
Given that  both the HM and the FKM realize phases with long range order,  we expect the AHM to also have such
phases. Based on the nature of the known solutions of the aforementioned models,  here, we assume  
a checkerboard-like order for the AHM, as well. This   implies the existence 
of two sublattices $A$ and $B$ such that each site of sublattice $A$ is connected only 
to the sites of sublattice $B$ and
viceversa. 
In the AHM, the existence of two inequivalent sites
necessitates a mapping onto two impurity models similar to the one described in (\ref{eq:impproblem}) and the
respective Green's functions obey the following self consistency conditions,
\begin{eqnarray}
\label{eq:ordersc}
\mathcal{G}_{0\sigma A}^{-1}(i\omega_n) &=&i\omega_n+\mu_\sigma -t_\sigma^2G_{\sigma B}(i\omega_n)
\nonumber \\
\mathcal{G}_{0\sigma B}^{-1}(i\omega_n)&=&i\omega_n+\mu_\sigma -t_\sigma^2G_{\sigma A}(i\omega_n)
\end{eqnarray}
\noindent
In the limits $r=0, 1$, this set of four equations  reduces to two. 
For the case of general asymmetry $(0\leq r<1)$, one needs to deal with the 
four different self-consistency conditions, which 
make the problem of the AHM considerably harder.
Technically, we have to simultaneously solve two different quantum impurity problems, 
with two different effective baths each.
Solutions with incommensurate  order  which might exist in this model, especially  away 
from half-filling\cite{incomhm,incomfk,camjayi} are not considered in this paper.
From the structure of the mean-field equations Eqs. \ref{eq:ordersc}, 
we find that for checkerboard-like order at half-filling, the Green's functions obey the following constraints:  
\begin{eqnarray} \label{eq:simp}
\rm{Re} G_{A\sigma}(i\omega_n)  &= &-\rm{Re} G_{B\sigma}(i\omega_n) \nonumber \\
 \rm{Im} G_{A\sigma}(i\omega_n) &=  &\rm{Im} G_{B\sigma}(i\omega_n)
\end{eqnarray}
\noindent
We term this symmetry broken state as one with  orbital order (OO), and, in the symmetric case for r=1 it coincides with the familiar N\'eel antiferromagnetic state.

From Eqs.\eqref{eq:ordersc} and \eqref{eq:simp}, we see that  unlike the simpler  case of the HM ($r=1$) , 
here $G_{Al}$ and  $G_{Bh}$  can be different for generic values of r$\ne$1.
This implies the possibility of a charge density wave (CDW) order, as we shall show below. 



\subsection{Long-range order phase}
\label{subsec:longrangequal}

Restricting ourselves to the half filled case with equal numbers of light and heavy particles,  we find that in the
orbitally ordered phase,  the average occupations of different orbitals on  the A and B sublattices satisfy the
constraint
$ n_{Al}+n_{Bl}=n_{Ah}+n_{Bh}=1$.
This implies that 
\begin{equation}
\label{eq:orbital_order}
 n_{Al}-n_{Ah}=-(n_{Bl}-n_{Bh})
\end{equation}
Additionally, the sublattice occupation  $n_\alpha=n_{\alpha l}+n_{\alpha h}$ ($\alpha=A,B$)
satisfies the relation
\begin{equation}
\label{eq:charge_order}
 -\left(1-n_{A}\right)=1-n_{B}
\end{equation}
\noindent
Eq. \eqref{eq:orbital_order} indicates that the orbital polarization in sublattice $A$ is the opposite to that of $B$. 
The order parameter can now be defined as \oop$=\langle n_{\alpha l}-n_{\alpha h}\rangle/2$, with, $\alpha$ being either $A$ or $B$. 
In the HM limit, this orbital order translates into a spin density wave with N\'eel antiferromagnetic order, and \oop\ is the 
familiar staggered magnetization.

\begin{figure}
\centering
\includegraphics[width=8cm,angle=0]{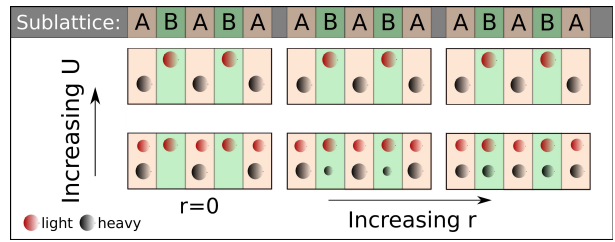}
\caption{\label{fig:ordercartoon}Sketch of  different  ordered solutions of the AHM. 
Black (lower) dots, represent particles in the heavy 
band, while red (upper) dots represent those in the light-one. Bigger dots represent higher occupation. In the SU(2) symmetric point 
$r\rightarrow1$ (right-most pannels), only orbital order is possible, and orbital polarization increases with $U$. 
As $r$ decreases, CDW ($n_A\not=n_B$) coexists with orbital order (lower left and central panel). CDW formation, nevertheless, decreases 
with increasing interactions (top panels), since in the Mott insulating state, double occupied sites are energetically unfavorable.}
\end{figure}

Note that Eq. \eqref{eq:charge_order} indicates that orbital ordering  coexists with a charge density wave formation 
in which one sublattice has increased its occupation with respect to the other.  
The  CDW  order parameter can be defined as \cdwop$=\langle n_{A}-n_{B}\rangle/2$.
We see that we have CDW order for all r$\ne$1.
In the limit of the HM,   both $l$ and $h$ orbitals are equivalent,   SU(2) symmetry is recovered and CDW 
order is  lost, since $n_{Al}=n_{Bh}$ and  orbital order translates to N\'eel antiferromagnetic order. 
An OO with a CDW, in contrast,  is realized in the FKM limit\cite{fk}. On heating, these ordered 
phases  survive upto  a certain critical temperature $T_o$, which coincides with the
N\'eel temperature in the HM, and with the $T_{\textrm{CDW}}$ in the FKM case.
To help visualize  the variety of states,  
we sketch in  Fig.~\ref{fig:ordercartoon}, the different phases as a function of the 
interactions and the hopping asymmetry.

\section{Results}
\begin{figure}
\centering
\includegraphics[width=8cm,angle=0]{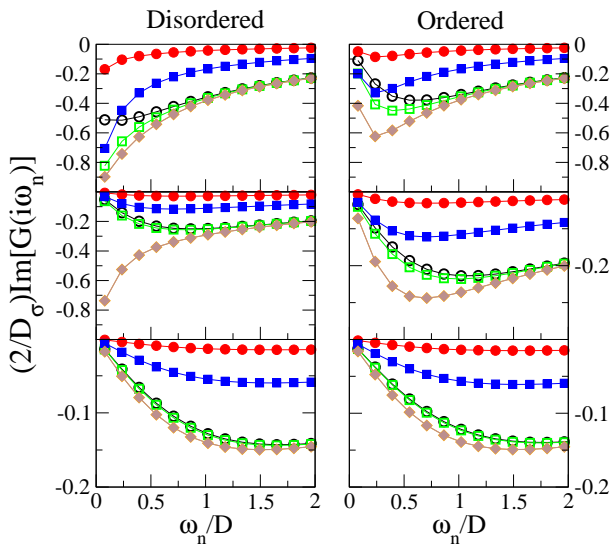}
\caption{\label{fig:greensfunc} $\textrm{Im}G(i\omega_n)$ for $T/D=0.025$. The left (right) column corresponds to  the disordered (ordered
) solution for $U=D,2D,3.5D$, for the first, second and third rows, respectively. Empty (filled) symbols correspond to the light (heavy)-orbital, while 
circles, squares and diamonds correspond to $r=0.1,0.4,1$, respectively. For the ordered state, note that as shown in Eq. \ref{eq:simp}, 
$\textrm{Im}G(i\omega_n)$ is independent of the sublattice.}
\end{figure}

We now solve the  relevant impurity problem  with the associated self-consistency condition in both
the ordered and the disordered phases, using
exact diagonalization and quantum Monte Carlo methods \cite{bible}.  
The use of both methods allows us to crosscheck the results and also obtain  
the finite temperature  phase diagram in the whole relevant parameter range of the model.
In Fig.~\ref{fig:greensfunc},  we  plot the finite-$T$ Matsubara local Green's functions  
for different  values of the parameters  $U$ and $r=$\ra\  for $T/D$=$D/\beta$=$0.025$. 
Although detailed features like the densities of states are technically difficult
to extract from the Matsubara  Green's functions, some basic properties can be readily obtained.
For instance,  the extrapolated value of $G(i\omega_n \to0)$  yields the density of states at the Fermi energy, which is
expected to be   finite
in  a metal, while it  goes to zero in an insulator at low temperatures. 
This qualitative feature  can be observed in the solutions   for the disordered (ordered)  phase  shown in the left (right) panels of Fig. \ref{fig:greensfunc}  for various values of $r$ and the  interaction $U$.
This feature allows us to pinpoint the precise
 location of the metal-insulator transition in the model. 
Moreover, within DMFT,  if the metallic solution is a normal Fermi liquid, 
 it can be shown that the density of states at the Fermi 
energy at $T=0$ for any $U$, remains pinned at the $U=0$ value \cite{bible}. 
 Consequently,  if the extrapolated value of  the Green's 
function does not go to zero (i.e., is not an insulator), but violates the 
pinning condition, then the  self-energy would be non-vanishing at $\omega_n \to 0$, 
hence the states have a finite lifetime, and the solution corresponds to a non-Fermi liquid metallic state. 
All of these features, coupled with the estimation of  various order parameters,  help us obtain
an interesting and rich finite temperature phase diagram for the AHM.

\subsection{Disordered phase}
\label{subsec:pmphase}
\subsubsection{Phase diagram}

\begin{figure}
\centering
\includegraphics[width=8cm,angle=0]{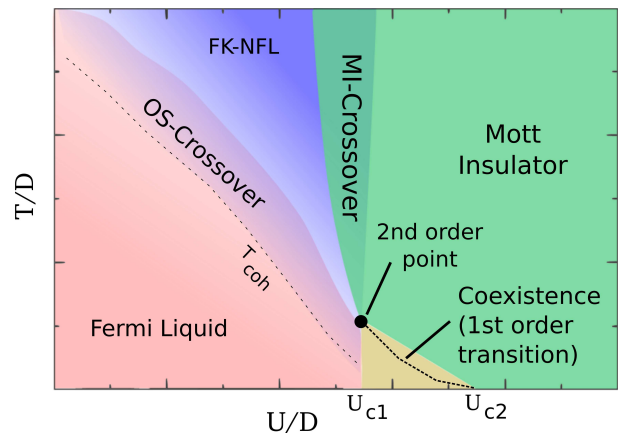}
\caption{\label{fig:tcoh}Schematic $T-U$ phase diagram for a generic value of the parameter $r$
\cite{emilio_prb}. 
At low $T$ there are three distinct 
regimes: a Fermi-liquid state with two different quasiparticles; a
coexistence zone, and a Mott insulating phase. In the metallic regime, above $T_{coh}$ there is a crossover to a Falicov-Kimball-like non-Fermi liquid state. 
Increasing interactions from this phase, the system crosses over to a bad insulating state (a finite $T$ Mott insulator).}
\end{figure}

The disordered phase of the AHM, has been discussed in detail in recent publications (Refs. \onlinecite{emilio_prb,Capone}). 
Here, for completeness and to facilitate the ensuing   discussion, we briefly summarize
the main features of the disordered phase diagram.    It is schematically shown in Fig. \ref{fig:tcoh} for a generic value of the mass asymmetry parameter $r$. This phase diagram would be  appropriate for a fully frustrated system.

At low T, there exists a metal-insulator transition from a metal to a Mott
insulator as one increases the repulsion $U$.
The metallic phase is characterized by a Fermi liquid for all $r \neq 0$ and generically has two kinds of quasiparticles
with different masses. 
The light particles are more strongly renormalized, since the screening potential produced by the heavy ones is less efficient due to its lower mobility.
Despite the different mass renormalization, both quasiparticles disappear (their mass renormalization diverges) at a single critical $U=U_{c2}$,  signaling the
metal-insulator transition. On the other hand, if one starts at high values of the intereaction $U$, well in the Mott state,
and progresively decreases the strength, the insulating gap decreases, and eventually closes. At that point there is
an insulator to metal transition, which is denoted as $U_{c1}$. 
In between these two critical values, one finds a coexistence
regime, with two solutions, one insulating and one metallic, as depicted in Fig. \ref{fig:tcoh}.
The coexistence region extends in temperature from $T$=0 up to finite temperatures. 
Both solutions are thermodynamically stable within the disordered phase, thus the metal-insulator transition is
first order at finite $T$ and culminates in a second order critical point,  similarly to the HM case\cite{bible}.
The critical values of the interactions and the $T$-value of the critical end point of this regime,
i.e., the area of coexistence, 
decrease with decreasing $r$.  

At lower $U$, the metallic Fermi liquid state survives upto a coherence temperature
$T_{coh}(U,r)$. Above  $T_{coh}$, the system crosses over to
an orbital selective metallic phase, which closely resembles the 
corresponding non-Fermi liquid state of the FKM model. This intermediate temperature regime, can be thought of
as if the hopping of the heavy particles were effectively renormalized to zero, hence becoming localized, 
while the light ones remain mobile.  Orbital selective crossovers between qualitatively different metallic phases, though commonly observed in real materials, such as high-Tc pnictides, heavy fermions, etc.,  are often not well  understood.   In this context, our model may be viewed 
as a minimal model for a non-Fermi liquid state emerging from an orbital-selective mechanism.

\begin{figure}
\centering
\includegraphics[width=8cm,angle=0]{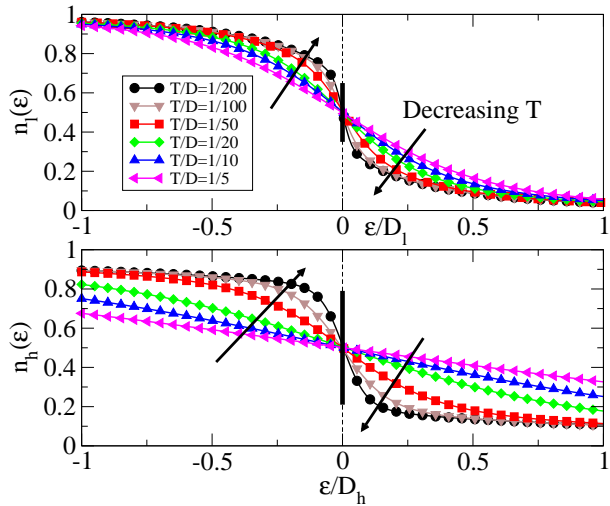}
\caption{\label{fig:nepsT}Density distributions of  the light and heavy orbitals for $r=0.2$ and $U/D=0.8$. 
As $T$ decreases, $n(\epsilon)$ tends to the usual Fermi-liquid distributions where there is 
a discontinuity at the Fermi energy (black arrows show the tendence of $n(\epsilon)$ as $T$ reduces). The vertical black line at $\epsilon=0$ shows the expected jump at $T=0$.}
\end{figure}

\subsubsection{Momentum dependent density of states}
Since the AHM is more generic than the Hubbard 
model, it may be easier to realize in cold fermionic mixtures.  Moreover, 
from the previous discussion it should be clear that the AHM presents the tantalizing possibility of 
realizing non-Fermi liquid states at relatively high temperatures.  These types of poor metallic 
states are central to many open problems in condensed matter physics. 
This  immediately raises the question of what would be
a measurable physical quantity, which is sensitive to the orbital selective crossover to the non-Fermi
liquid state. 

One practical probe  would be  time-of-flight experiments \cite{rmpcoldatoms}, 
where the momentum resolved density distribution
$n(k)$ is measured. In a conventional Fermi-liquid at zero temperature, $n(k)$ 
has a jump of size $Z\leq 1$ at the Fermi momentum $k_F$, with $Z$ being the
quasiparticle residue. 
Replacing the momentum  by the single particle energy $\epsilon$, the  
resulting quantity $n(\epsilon)$, which is more easily obtained within DMFT, would also exhibit 
the same jump at the Fermi energy $\epsilon=\epsilon_F$. At nonzero  $T$,  the discontinuity in $n$ is smeared 
within an energy scale given by $T$ (see Fig. \ref{fig:nepsT}). 
In a non-Fermi liquid state, $Z=0$, and there is no discontinuity even at $T=0$.

\begin{figure}
\centering
\includegraphics[width=8cm,angle=0]{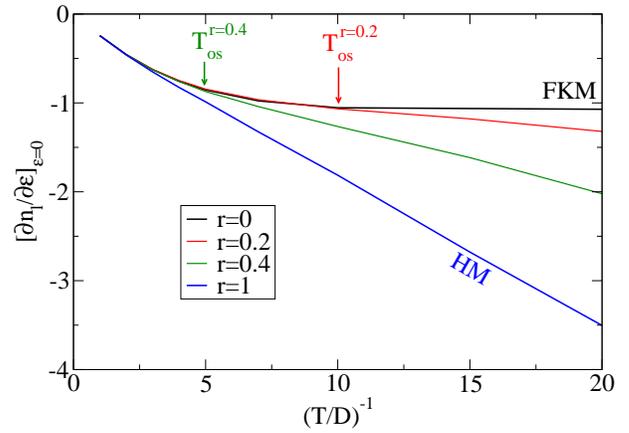}
\caption{\label{fig:slope} 
Slope of the light particles density distribution as a function of $(T/D)^{-1}$, for $U/D=0.8$. 
For $T>T_{\rm OS}$, the metallic behavior
of the AHM, coincides with the non-Fermi liquid metallic state of the FKM. For this value of $U$, 
$T_{\rm OS}(r=0.4) \simeq 0.2D$ and $T_{\rm OS}(r=0.2) \simeq 0.1D$.}
\end{figure}

Since the crossover of the AHM is at finite $T$, we do not expect to see sharp changes  in the density distribution. 
Nonetheless, going from the HM towards the FKM (reducing \ra\ and keeping all other variables constant), 
$n(\epsilon)$ 
evolves to $n^{\textrm{FKM}}(\epsilon)$, 
for $r<r_{crit}$.
However, we find that the slope of 
$n_l(\epsilon)$ at the Fermi energy, as a function of $T^{-1}$ is a better magnitude to pinpoint this crossover.  We plot the slope  in 
Fig.~\ref{fig:slope}  and  we clearly see that above a  certain temperature $T_{\rm OS}$, which depends on $r$ 
and $U$, 
$\frac{\partial n^{\textrm{AHM}}}{\partial\epsilon}$ approaches $\frac{\partial n^{\textrm{FKM}}}{\partial\epsilon}$. 
Therefore, this signals that the light orbital of the AHM  evolves to  the non-Fermi liquid state of the FKM  once 
orbital-selective crossover has been achieved.

\subsection{Ordered phase}
\label{subsec:orderphase}
In the non-frustrated case,   we expect the system to order at low temperatures.  Solving the equations discussed
in section \ref{subsec:longrange}, we obtain the 
 critical temperature $T_o(r,U)$,  below which the orbital order sets in, 
and the behavior of the order parameters $m_o$ and $m_c$, defined before.

\subsubsection{Ordering Temperature $T_o$}
\label{subsubsec:to}

\begin{figure}
\centering
\includegraphics[width=8cm,angle=0]{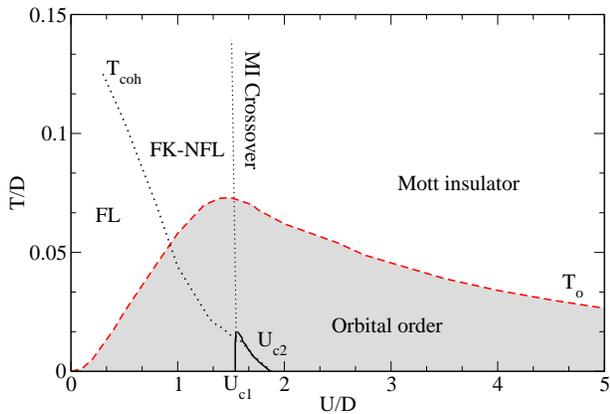}
\caption{\label{fig:phasediagram}$T$-$U$ phase diagram for $r=0.4$. $T_o$ (red dashed line) is the critical temperature. Orbital ordering shadows the metal-insulator transition, which occurs at $T<T_o$,  like in 
 the case of the HM\cite{bible}. 
The different phases of the disordered solutions are described in Fig. \ref{fig:tcoh}}
\end{figure}

We find that   analogous to the HM and FKM,   at $T=0$ and  for all $U\neq 0$,
the system  is an  insulator with long-range order which vanishes
above a critical temperature $T_o (U,r)$. 
The results are shown in Fig.~\ref{fig:phasediagram} for a generic value of $r$=0.4.
For comparison, in the figure we also show the main features of the phase diagram
discussed before, in Fig.~\ref{fig:tcoh}.
We observe that the metal-insulator transition, described in section \ref{subsec:pmphase} is shadowed by the OO 
phase, which is  thermodynamically more stable.  This behaviour is similar to the one realized in the 
Hubbard model \cite{bible}. 
It is interesting to observe that both the coexistence region, the orbital-selective thermal crossover and
the maximum in the orbital ordering temperature $T_o$, all occur at the same parameter region, where
the interaction $U$ is comparable to the model bandwidth $2D$ and the correlation effects are strongest.

\begin{figure}
\centering
\includegraphics[width=8cm,angle=0]{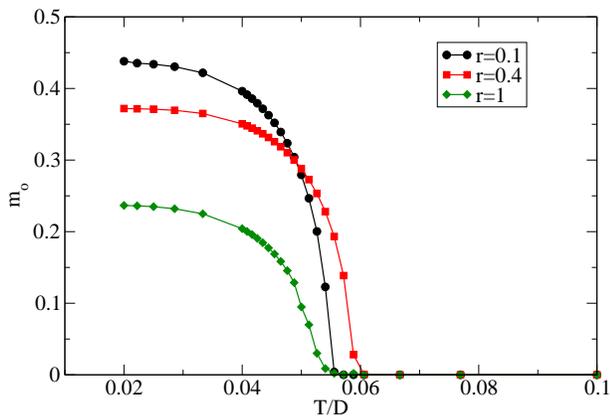}
\caption{\label{fig:mvsT}Orbital order parameter \oop\ for different values of the asymmetry $r$ and $U=D$.}
\end{figure}

\begin{figure}
\centering
\includegraphics[width=8cm,angle=0]{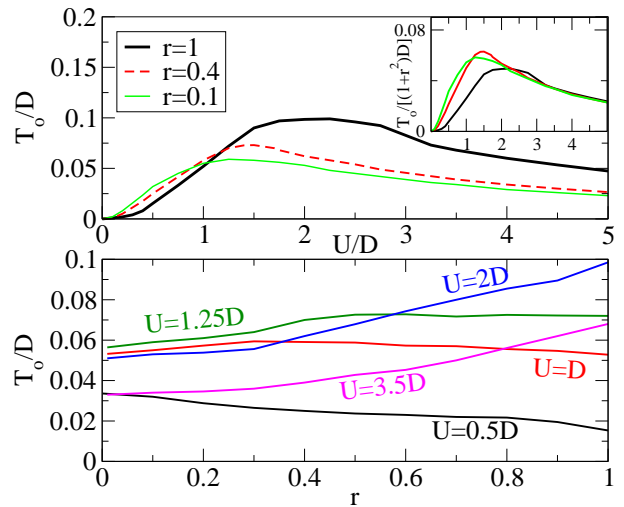}
\caption{\label{fig:tneel}Critical temperature $T_o$ as a function of interaction $U$ (upper panel) and
hopping ratio $r$ (bottom panel). The scaling $T_o\propto (1+r^2)$ at strong coupling, is shown in the inset of the upper panel.}
\end{figure}

Within DMFT, the simplest procedure to compute the critical temperature $T_o$ is to start at low $T$, where the 
converged DMFT solution is found to have orbital long-range order 
(i.e., \oop$\not=0$, see eq.~\ref{eq:orbital_order}) and then slowly increase the temperature  until the order vanishes
at some $T=T_o(r,U)$. We note that these transitions though continuous and second order, 
are rather sharp as can be seen  in Fig. \ref{fig:mvsT}, that shows the order parameter \oop($T$), 
for $U=D$ and several values of the hopping ratio $r$. 
In Fig. \ref{fig:tneel}, we plot the ordering temperature as a function of both, the 
interaction $U$ and hopping asymmetry $r$.

At weak coupling i.e., small $U$ and for all $r\neq 0$, we see an exponential behaviour , $T_o \propto \exp (- c/U)$, where $c$ is some constant, similar to the  Hartree mean field result for the  Hubbard model\cite{vdongen}.  
In the FK limit, we expect\cite{vdongenfk} $T_o \propto (U/D)^2\ln (D/U)$.
The qualitative dependence of $T_o$ on
$r$ can be understood by considering the competition between kinetic and Coulomb energy. In the disordered metallic region, when $U=0$, $T_o$ goes to zero, and there is no order. 
In the HM, as $U$ increases,  the ratio between potential and kinetic energy ($E_p/E_k$) increases, and so does $T_o.$\cite{bible}  Since  a  decreasing $r$  decreases $E_k$,  the ratio $E_p/E_k$  
and hence $T_o$ increase as $r$ decreases   in  accordance with Fig. \ref{fig:tneel}.

 To understand the non-monotonic behaviour of $T_o$  seen in  the Mott insulating side of the 
phase diagram, it  is useful to consider  the large-$U$ limit,  where  the AHM can be mapped\cite{fath} onto a {\it spin}-$1/2$ anisotropic 
Heisenberg (XXZ)  model described by the effective hamiltonian,
\begin{equation}
H_{spin}=J\sum_{<ij>}[\sigma^x_i\sigma^x_j+\sigma^y_i\sigma^y_{j}+\gamma 
(\sigma^z_i\sigma^z_{j}-1)]
\end{equation}
\noindent
 with  exchange constants $J=$\tup\tdw$/U$ and $\gamma=\frac{t_l^2+t_h^2}{2t_h t_l}$.
This model interpolates between the isotropic Heisenberg model 
at $r=1$ and the Ising model at $r=0$.  A mean field study of this spin model shows us that
the ordering temperature  $T_o$ in the large $U$ limit is 
proportional to $J\gamma \propto (1+r^2)$ consistent with the DMFT results, which show that  $T_o$ decreases 
with decreasing $r$ in the large $U$ limit (see inset of Fig.~\ref{fig:tneel}). This feature also persists in the intermediate $U$ regime as shown
in the lower panel of Fig.~\ref{fig:tneel}.   
Therefore, from the previous discussion, the 
non-monotonic behaviour of $T_o$ seen in Fig.~\ref{fig:tneel} is a consequence of the
increasing dependence of the ordering $T$ with $U$ at weak coupling, on one hand;
and of the decreasing dependence at strong coupling due to the super-exchange interaction, on the other.

\subsubsection{Orbital and charge order}
\label{subsubsec:orb-chorder}

\begin{figure}
\centering
\includegraphics[width=8cm,angle=0]{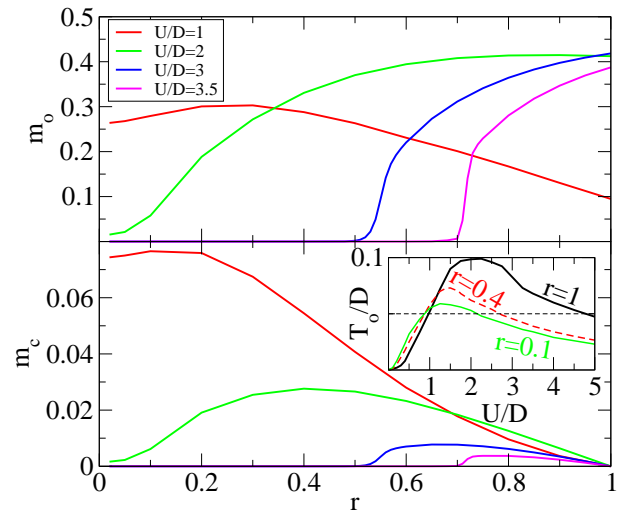}
\caption{\label{fig:op}Orbital order parameter (upper panel) and charge density wave order parameter (bottom panel) as a function of $r$ for  different values of $U$ and $T/D=0.05$. Inset: $T_o$ for different values of $r$ as a function of $U$. The horizontal dashed line corresponds to $T/D=0.05$, which is the 
temperature used for computing \oop\ and \cdwop.}
\end{figure}

We now discuss the orbital and CDW ordering seen in the ordered phase for $T<T_o$.
In Fig. \ref{fig:op}, we plot the order parameter \oop\ 
(defined in sec. \ref{subsec:longrangequal}) as a function of the asymmetry $r$ for different 
values of the interaction, at a fixed finite low $T$. 
The variation of \oop\ with $r$ is rather complex and depends on the value of $U$.
For   large enough  $U $, we observe a sudden increase of the orbital order parameter with increasing $r$.
This is because the energy of the OO state at large $U$, where we have a Mott insulator, is given
by the super-exchange, which increases with $r$, as discussed above.
At lower values of $U$, in contrast, the system is in a more itinerant state, a Slater like OO. Thus, increasing $r$
the system is driven by kinetic energy towards a metallic state, lowering the magnitude of the order parameter.
From a practical standpoint,
these results suggest that in  experiments,  for any given value of the interactions $U$, 
the phase transition  to the OO phase can be 
attained  by tuning either $T$ or $r$.  This second route to achieve the
long-range ordered phase  can be of particular interest in  cold atom systems  where it is often very difficult to  reach  the  low temperatures  associated with long range ordering. 

Another interesting aspect of the ordered phase, is that orbital order coexists with a genuine charge density
wave (CDW) state. This is illustrated in the bottom panel of Fig. \ref{fig:op} where the CDW order 
parameter \cdwop\ 
(defined in sec. \ref{subsec:longrangequal}) is plotted as a function of the hopping asymmetry for different 
values of the interaction.
We observe that the total number of particles per site is different depending on whether the site belongs to the A or B sublattice, 
as qualitatively depicted in Fig. \ref{fig:ordercartoon}.  The behavior of  \cdwop\ is non-monotonic, large 
at low interactions and small $r$, and smaller as one approaches the Hubbard limit, where it eventually vanishes.
This can be understood from the fact that
 large   $U$ is detrimental for  charge inhomogeneity, and penalizes the formation of a  CDW. 
Alternatively, an increase of 
$r$ also favors a more homogeneous state, reducing the order parameter $m_c$.
The situation at low $U$ is qualitatively different and  can be understood  using the  same arguments presented earlier for the behavior of $m_o$.

\begin{figure}
\centering
\includegraphics[width=8cm,angle=0]{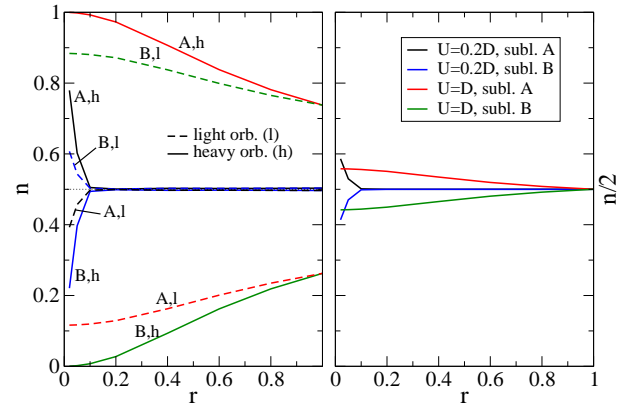}
\caption{\label{fig:nABsig}Left panel: Number of light and heavy particles on the A and B sites  as a function of $r$ for different values of  $U$ and $T/D=0.01$. Right panel: $n_A$ and $n_B$. Notice that $n_{A} \neq n_{B}$ for $U$=$D$, while for $U$=0.2$D$ the CDW disappears for $r>0.1D$.}
\end{figure}

Further  insight is gained by computing the contribution of each orbital to the formation of the CDW.
In Fig. \ref{fig:nABsig} we plot the mean number of particles in each of the two orbitals on sites of the 
A and B sublattices. We find that the occupation of the heavy orbital has a stronger staggered 
character than the light one ($|n_{Ah}-n_{Bh}|>|n_{Al}-n_{Bl}|$). This can be understood by a continuity argument:   in the FKM limit ($r$=0) and low U, the heavy 
orbital is localized while the light one remains itinerant.
For higher values of $U$, where one is within the 
ordered phase for all $r$, we see that this pronounced staggering persists, 
though becomes weaker as $r$ increases and eventually
vanishes at the Hubbard point ($r$=1).


\begin{figure}
\centering
\includegraphics[width=8cm,angle=0]{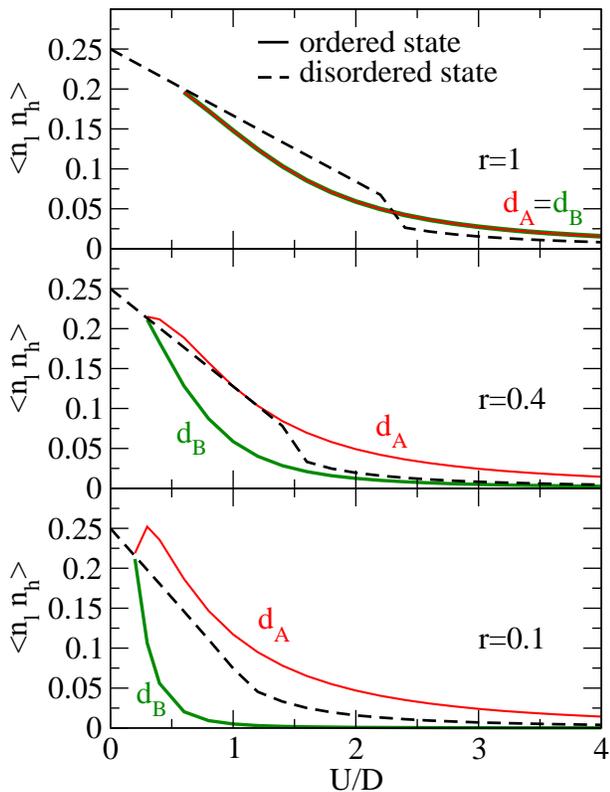}
\caption{\label{fig:doccAB} Double occupancy $d_A$ and $d_B$ (solid lines) for different values of $U$  and  $r$ at $T/D=0.01$. For comparison,  the dashed lines represent the equivalent curves for the disordered state. The sharp behavior at low $U$ is a signature of the disorder-order transition at finite $T$.}
\end{figure}

All of these results are consistent with the behaviour of the double occupancy, which can actually be experimentaly measured in  cold atom systems\cite{docc}.
In Fig. \ref{fig:doccAB}, we plot the double occupancy $d=$\docc\ in both sublattices
in the ordered phase and also its value in the disordered phase. 
The fact that true CDW order accompanies the orbital order immediately implies that $d$ can show different
behaviours on each of the sublattices. This is indeed  the case as shown in Fig. \ref{fig:doccAB}, where $d_A$ and $d_B$ 
are different within the ordered phase ($T < T_o$), when $r\not=1$.

In the symmetric case, $r=1$, we observe that, for moderate to strong interactions, 
the double occupancy of the ordered phase is higher than in the 
disordered one. This
is due to the fact that in a state with staggered order, hopping to a neighboring  site 
is less penalized by Pauli blocking  with respect to the situation in a disordered state\cite{prl105_germans}.
On the contrary, for small to moderate coupling the behavior is the inverse, since in the disordered metallic state 
the particles delocalize on both orbitals, increasing the probability of double occupation. The ordered state remains
an insulator even at small $U$, thus favoring orbital localization. 
At low enough $U$, the ordered state eventually becomes unstable at the finite $T$ of 
our numerical calculations, ie, when $T>T_o(U)$.

For $r \neq 1$, we observe the effects of the onset of the CDW. 
The behavior of the double occupancy becomes different on each sublattice, as previously discussed. 
At low couplings, we find that the difference has a sudden increase, which reflects the steep rise of the 
orbital order-parameter \oop\ as the system enters the ordered phase. This shows that an abrupt change in the sublattice double occupation is a good probe 
of CDW order in a system.
This sharp behaviour reduces with increasing $r$ as expected because the CDW order decreases with increasing
$r$, fully disappearing in the limit of $r=1$. We also see that, for any value of $U$, the double occupancy is higher
in the HM and decreases with decreasing $r$ which is consistent with the fact that as $r$ decreases it gets harder
for the heavier particle to hop from one lattice site to another.
These results confirm  that asymmetric hopping amplitudes favor a CDW, which
can be experimentally detected. Alternatively, this also suggests that the density distribution can be manipulated by tuning the hopping ratio $r$.

\subsubsection{Thermodynamics}
\label{subsubsec:thermodynamics}

One of the challenging aspects of cold atomic systems in optical lattices is the determination of
their temperature. To this end, it is important to know the thermodynamic properties of the system.
With this issue in mind,  
we focus on the computation of  the fundamental thermodynamic quantities: specific heat and
entropy of the AHM in this section. We set the Boltzmann constant $k_B=1$.

\begin{figure}
\centering
\includegraphics[width=8cm,angle=0]{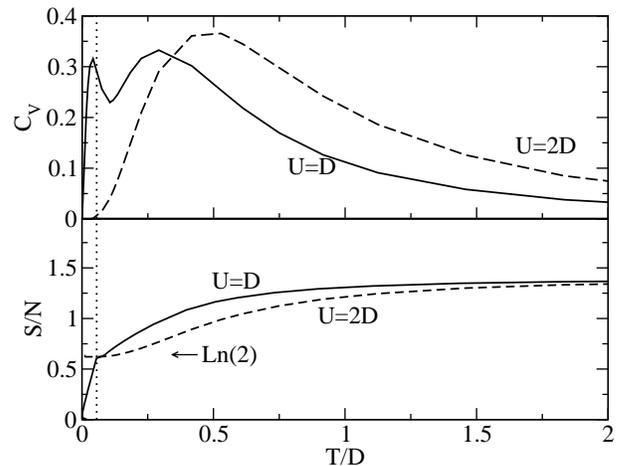}
\caption{\label{fig:cvandentropy-dis}Specific heat and entropy per site 
as a function of $T$ for $r=0.4$, in the disordered case for   $U=D$ (metal)  and $U=2D$ (insulator). 
For reference, $T_{coh}$ is indicated by the vertical dotted line. It corresponds
to the destruction of the Kondo effect, the first peak in the \cv\, and the $\ln(2)$ plateau in $S/N$.} 
\end{figure}

In our previous paper \cite{emilio_prb}, we discussed the specific heat \cv\ and the entropy 
per site (or equivalently per particle if working at $n_{tot}=1$) $S/N$=$s$ in the disordered phase of the AHM. 
The main features can be summarized as follows (see Fig. \ref{fig:cvandentropy-dis} where we plot these
quantities at a finite generic value of $r$): 
In the metallic phase at low $U$, 
the \cv\ and $s$ are linear in $T$, for all values of the asymmetry parameter $r\not=0$ and 
 $T < T_{coh}$. This behavior is a signature of the Fermi liquid groundstate.
 At $T \approx T_{coh}$, the \cv\ develops a peak and the entropy has a small plateau at $s=\ln(2)$, that corresponds
to the two orbital degrees of freedom. In the Kondo impurity problem, this scale indicates
the  destruction of  the Kondo singlet state [$(|lh\rangle - |hl\rangle) \to (|l\rangle \bigotimes |h\rangle)$] .   
Increasing $T$ further, $s$ increases  and eventually saturates at 
$s=\ln(4)$ at high $T$. This value corresponds to the four different states available per site, 
ie $|0\rangle,|h\rangle,|l\rangle,|lh\rangle$. On the other hand, 
 \cv\ develops a second peak due to the excitation of states of the incoherent Hubbard bands.

The  behavior of the entropy in the 
Mott state of the AHM, shown in Fig.~\ref{fig:cvandentropy-dis}, resembles that of 
the metallic state of the standard FKM\cite{emilio_prb}. However, the similarity is superficial, and
is due to different physical reasons. In the latter, the heavy particles have zero hopping amplitude and are localized, 
thus   there is an equal probability of finding a heavy particle or not on any given site. Therefore, there
are two occupation states associated with any site and this
contributes a factor of $\ln(2)$ to the entropy, even as $T \to 0$. As $T$ increases, the entropy  increases monotonically  to the asymptotic value $\ln(4)$. 
However, in the Mott insulating state of the AHM, 
 both particle species are localized and, in contrast to the FKM case that we just described, 
 there are two different ways of singly occupying
a site ($|h\rangle$ or $|l\rangle$), hence the $\ln(2)$ in the entropy.
As before, at high $T$, density fluctuations are thermally 
activated, hence the entropy and $s$ also goes continuously to $\ln(4)$. 
Regarding the behavior of the \cv\, since the insulating Mott state of the AHM is characterized by the 
opening of a correlation gap, the specific heat displays activated behavior at low $T$.

\begin{figure}
\centering
\includegraphics[width=8cm,angle=0]{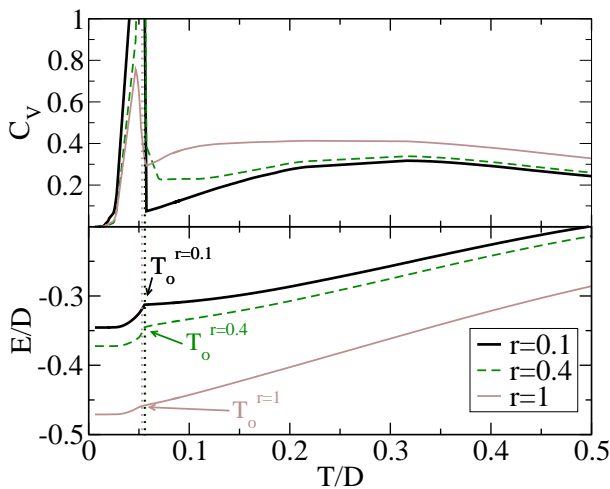}
\caption{\label{fig:cvU1order}Specific heat and internal energy as function of $T$ for $U=D$, and different values of $r$ in the ordered phase. 
The first peak in  \cv\ reveals  a formation of an orbitally ordered phase, while the second, the localization of the particles. The sharp reduction of
the internal energy when decreasing $T$, reveals the formation of an insulated-orbitally ordered state, at $T_o$. 
For $U=D$, $T_o$ is weakly dependent on $r$, as shown in Fig. \ref{fig:tneel}.}
\end{figure}

\begin{figure}
\centering
\includegraphics[width=8cm,angle=0]{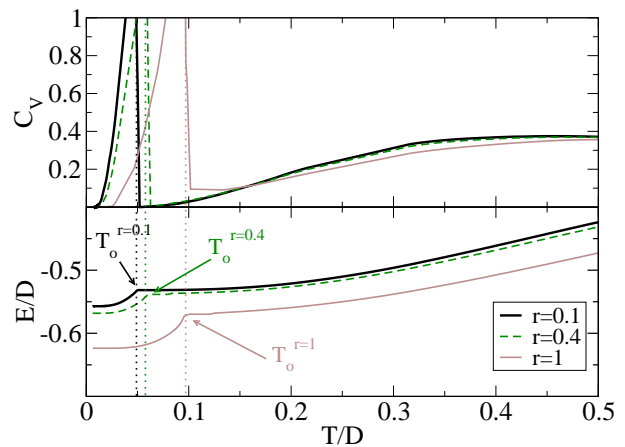}
\caption{\label{fig:cvU2order}Specific heat and internal energy as function of $T$ for $U=2D$, and different values of $r$ in the ordered phase.  Note that for $U=2D$,
$T_o$ increases with $r$.}
\end{figure}

The behavior of the specific heat in the ordered phase shows clear signatures of a phase transition.
The results are shown in Figs. \ref{fig:cvU1order} and \ref{fig:cvU2order} for characteristic 
values of the parameters $r$ and $U$.
At very low $T$, the \cv\ has activated behavior (the internal energy is almost constant), due to the opening of 
 the insulating gap  in the orbitally ordered state. 
Increasing $T$, the system eventually reaches the transition temperature $T=T_o$  where the order is lost.
The internal energy plotted in  Fig. \ref{fig:cvU1order} shows a sudden 
upturn as the state becomes unstable. At the critical point there is a kink
which translates to a divergence in the specific heat. 

\begin{figure}
\centering
\includegraphics[width=8cm,angle=0]{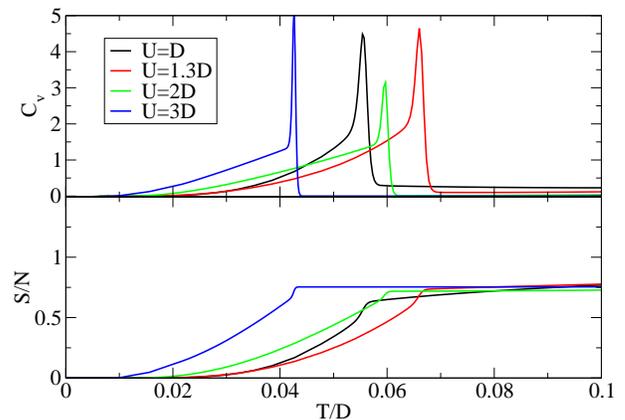}
\caption{\label{fig:entropyorder}Specific heat and entropy as a function of $T$ for $r=0.4$ and $U/D=1,1.3,2,3$, in the ordered phase. Above $T_o(U)$, where the $C_v$ has a kink, the entropy coincides with that of the disordered state. For low enough $T$, the activated behavior is non-monotonic in $U$, revealing the different natures of the ordered state with varying $U$.}
\end{figure}

In Fig. \ref{fig:entropyorder} we plot the entropy in the ordered state 
for $r=0.4$ and several values of $U$.
The data shows activated behavior at very low $T$ and a rapid increase
of the thermal excitations as the temperature approaches $T_o$ and the ordered
moment melts (see Fig.~\ref{fig:mvsT}). For $T$ above $T_o$, in the disordered phase,
the entropy is identical to that of Fig. \ref{fig:cvandentropy-dis}. 
The non-monotonic behavior of the position of the maxima, is simply a consequence of the
peculiar behaviour of $T_o$ with $U$ that was already discussed Sec.~\ref{subsubsec:to}.


\subsection{Proposal for cooling mechanism: Entropic chromatography}
\label{subsec:cooling}

Despite the fact that experiments on cold atoms in optical lattices are done at ultra low temperatures, it is necessary to achieve even lower temperatures, 
to access new interesting strongly correlated quantum phases.

One of the theoretical schemes proposed for cooling, is based on the idea of
embedding a fermionic gas in an optical lattice within a bosonic bath, and then `squeezing' out the 
entropy of the fermions.
This may be achieved by first shrinking their trap, thus compressing the fermionic gas towards a band insulator state. 
This state has no free degrees of freedom left, hence, no entropy (with the exception of the states at the edges of the trap). 
The lost fermionic entropy is thus transferred to the bosonic bath, and the bosons are then released.  The remaining fermions
are in a low entropy state, therefore at a much lower temperature \cite{Ho}.
Another proposal\cite{Bernier}, suggested  modifications to the shape of the harmonic trap 
so as to create a deep dimple in the center where one would form a band insulator. 
This separates the system into an entropy poor state at the center, surrounded by 
entropy rich regions. The entropy rich particles could subsequently be removed from the system by 
partially opening the trap and  the remnant band insulator could then  be 
adiabatically relaxed to the relevant experimental regime.

Motivated by the present study of the AHM, 
here, we propose an alternative scheme to cool the system based on the principal feature of the AHM i.e., the
hopping asymmetry.  In 
Fig.\ref{fig:entropy-nonint},  we plot the entropy per site $s$, in the non-interacting limit ($U=0$) limit of the AHM at half
filling.  We see that most of the entropy is carried by the heavier particles for $T  < 0.4 D$.  The difference in
entropy between the two species is more pronounced at very low temperatures.  Though we have not included the
effect of the trapping potential in our calculation, we do not expect 
it to significatly change the nature and validity of the cooling mechanism that we shall describe.

The ability to concentrate most of the entropy in the heavy species by manipulating the hopping amplitudes
of one species,
immediately suggests the possibility of  an entropic chromatography method.  One can repeatedly
concentrate entropy in one of the species and use the other as a thermal bath, which is then partially evaporated,
liberating entropy (ie heat) at each successive step.
This leads to  a big reduction of  entropy  of the remanent system, 
hence reducing its temperature.  Our cooling
scheme is based on this entropic chromatography, supplemented with a series of
system parameters modifications,  to ensure that  the physical system stays within the required
filling regime (ie one particle per site on average). 
This  scheme is similar to those mentioned above, in that we separate the system into
entropy rich and entropy poor sectors, with the main difference being that here the entropic separation is not spatial.   
The advantage of this procedure is that it does not require an additional  Bose condensate to act as a reservoir 
or additional lasers to modify the trapping potentials in a precise manner.
Our scheme essentially relies on the possibility of having two different fermionic atomic species within
their respective atomic traps, which can be independently adjusted.
We describe our cooling procedure in detail below.

We  consider a system  where the light and heavy particles have their respective trapping
potentials and their interaction   is described by the AHM. 
We limit our analysis to the disordered phase of the AHM. However, 
the extension of the idea to the long-range ordered phase can  be performed along similar lines.
In this section, for  the sake of clarity in the description of the procedure,  we denote the two atomic species as
particles of type 1, and 
type 2. We assume  that all the system parameters (hopping, $U$, trapping potential) can be tuned adiabatically and that the system thermalizes to an equilibrium state.  The cooling scheme to be described below is useful and applicable provided
the initial temperature and entropy of the system  is such that in the non-interacting limit, the difference in the entropies of
the light and heavy particles is substantial. 


We  summarize the cooling scheme in the following steps :

\begin{enumerate}[i)]
\item Start by loading $N_1=N_2=N$ particles of each fermionic species into the optical lattice  at finite $U$ and hoppings
\label{enum:firststep}
\begin{eqnarray}
t_1&=&t\nonumber\\
t_2&=&\alpha t
\end{eqnarray}
 with an  asymmetry $\alpha \ll 1$. The parameter $\alpha$ actually plays the same role as the ratio $r$, however,
we use this convention to emphasize that $\alpha$ is varied during the cooling procedure.  
The entropy of this initial state  is denoted as $S_{tot}^{(i)}$. The trap and the optical lattices
 are adjusted such that the system is half-filled.
\item  The interaction is now adiabatically tuned to $U=0$.  Since this is an isentropic process, the entropy of the resulting system is the same as that of the initial state. However, as shown in Fig. \ref{fig:entropy-nonint}, in this non-interacting limit, the entropy of the system is now mostly carried by the heavy particles $2$. 
\label{enum:secondstep}

\begin{figure}
\centering
\includegraphics[width=8cm,angle=0]{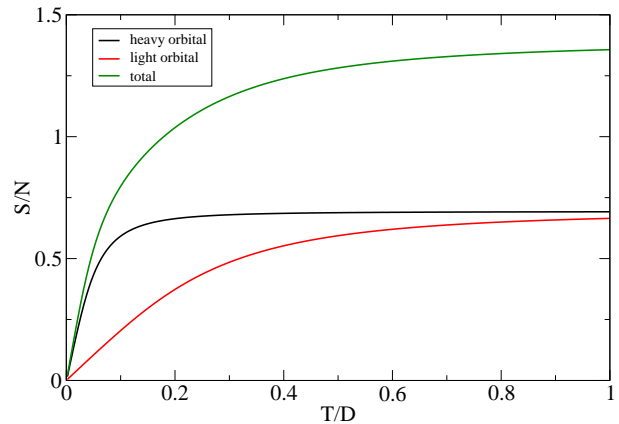}
\caption{\label{fig:entropy-nonint} Entropy per lattice's site in the non-interacting system for $r=0.2$.} 
\end{figure}

For non-interacting fermions (or quasiparticles in a Fermi liquid), at low $T$ the entropy  is proportional to $T$ \cite{noziere}:
$S = {\kappa_B^2k_FmT}/{3\hbar}$
where  the  mass of the particles (or the effective mass of the quasiparticles in the 
Fermi-liquid regime)  $m$ is proportional to the inverse of the
hopping amplitude $t$ and $\kappa_B$ is the Boltzmann's constant.  This implies that 
at this step  of the cooling process, the entropy of the  system   can be expressed as: 
\begin{eqnarray}\label{eq:ent-in}
S_{tot}^{(i)}&=& N s_1^{(1)}+Ns_2^{(1)}= \frac{ \kappa NT^{(1)}}{t}+\frac{\kappa NT^{(1)}}{\alpha t}\nonumber\\
&=&\kappa N\frac{T^{(1)}}{t}\left(1+\frac{1}{\alpha}\right)
\end{eqnarray}
where  $s_j^{(1)}$ is the entropy per particle of species $j$=1, 2 at  this step,  and $T^{(1)}$ is the 
new temperature resulting from the adiabatic variation of $U$. $\kappa$ is some proportionality constant which is
the same for both species (cf. Fig.\ref{fig:entropy-nonint}).

\item  Open the trap of particles $2$, and free half of them. The released entropy ($S_{-}^{(1)}$) is:
\label{enum:free2}
\begin{eqnarray}
S_{-}^{(1)}=\frac{\kappa NT^{(1)}}{2\alpha t}
\end{eqnarray}

\item Turn on $U$ adiabatically so the particles interact and thermalize (we assume  here
that the relaxation time is fast). The system now has an unequal number of particles 
of both species and a total entropy $S_{tot}^{(2)}= S_{tot}^{(i)} - S_-^{(1)}$

\item To eliminate the excess number of particles of species 1 and to reduce the entropy further, we  adiabatically invert the ratio $t_2/t_1$, in order to have,
\begin{eqnarray}
t_1 &=&\alpha t\nonumber\\
t_2&=&t
\end{eqnarray}

\item  The interaction is again adiabatically tuned to $U=0$.  We  now have a state with total entropy $S_{tot}^{(2)}$ such that
\begin{eqnarray}\label{eq:stot2}
S_{tot}^{(2)}&\equiv& N s_1^{(2)}+\frac{N}{2}s_2^{(2)}=\frac{\kappa NT^{(2)}}{\alpha t}+\frac{\kappa NT^{(2)}}{2t}
\end{eqnarray}
where $s_1^{(2)}$ and $s_2^{(2)}$ denote the entropies per particle of the two species  and $T^{(2)}$ the resulting
temperature at this step.

\item We  open the trap for particles $1$, and  eliminate $N/2$ particles of type 1. The freed entropy at this step is
\label{enum:free1}
$S_{-}^{(2)}=\frac{\kappa NT^{(2)}}{2\alpha t}$
and the total residual entropy is 
\begin{eqnarray}\label{eq:ent-3}
S_{tot}^{(3)}= S_{tot}^{(2)} -S_{-}^{(2)}
\end{eqnarray}

\item To obtain the final state with $N/2$ particles of each species and to thermalize the system so as to have
a well defined temperature, we, once again, invert the ratio $t_2/t_1$, at finite $U$. 
The traps of both species are narrowed, adapting the number of sites  ensuring that
 we have a half-filled system.
\label{enum:last}

\item In order to compare the temperature of this final  state with the initial one, we turn off $U$. 
This  final state  has equal numbers of particles $N/2$ of each species  and the  same hopping ratio  as in step \ref{enum:firststep} and an entropy 
\begin{eqnarray}
\label{eq:Sfin}
S_{tot}^{(f)}=S_{tot}^{(3)}=\frac{\kappa NT^{(f)}}{2t}(1/\alpha+1)
\end{eqnarray}
The  isentropic processes also ensure that the temperature of this state $T^{(f)}=T^{(2)}$,
which follows from Eqs.~(\ref{eq:stot2}), (\ref{eq:ent-3}) and (\ref{eq:Sfin}).
Then, by matching the final entropy of Eq. \ref{eq:Sfin} with
\begin{eqnarray}\label{eq:Sfin2}
S_{tot}^{(f)}= S_{tot}^{(1)} -S_{-}^{(2)}-S_{-}^{(1)}
\end{eqnarray}
 we find that
 \begin{eqnarray}\label{eq:temp}
\frac{T^{(f)}}{T^{(1)}}=\frac{2\alpha+1}{\alpha+2}
\end{eqnarray}
\end{enumerate}
Note that the initial temperature of the mixture $T^{(i)}$ is not know, the first temperature that could be defined
was $T^{(1)}$ at step (ii) with Eq.~\ref{eq:ent-in}.
From the last expression and Eqs.~\ref{eq:ent-in} and \ref{eq:Sfin} it also follows that the 
ratio of the total final and initial entropies is 
$S_{tot}^{(f)}/S_{tot}^{(i)} = T^{(f)}/2T^{(1)}$, thus that  the entropy per particle in the $U=0$ limit can be reduced
by the same factor $T^{(f)}/T^{(1)}$.  We also see from (\ref{eq:temp}), that no reduction of 
entropy or temperature is achieved
in the Hubbard limit $\alpha \to 1$  and that the cooling is entirely a consequence of the anisotropy in the hoppings.  
The maximal reduction  $T^{(f)}= T^{(1)}/2$ is obtained in the FK limit, $\alpha\rightarrow0$.  
Note that depending on the number of particles loaded into the optical lattice and the final system
required, the cooling procedure can be repeated a finite number of times, resulting in a drastic reduction of the temperature.

However, we should mention that
in cold atom experiments, currently accessible entropies per particle are of the order of $s \simeq 1.5$. 
This value is well outside
the linear regime discussed here (cf. Fig. \ref{fig:entropy-nonint}) and the analytical result obtained here,
though helpful to illustrate the mechanism, is quantitatively invalid. Nevertheless,
we emphasize that the cooling procedure 
 \ref{enum:firststep}-\ref{enum:last}, should still qualitatively work within a wider 
window of $T$,  since the large difference in entropy persists well beyond the
linear regime (cf. Fig.~\ref{fig:entropy-nonint}).

\begin{figure}
\centering
\includegraphics[width=8cm,angle=0]{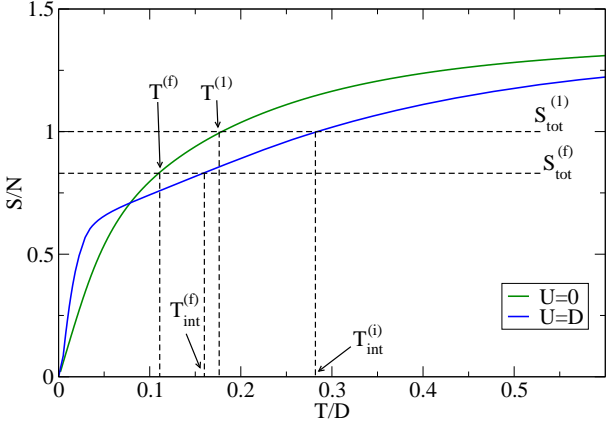}
\caption{\label{fig:entropycooling}Entropy as a function of $T$ with a hopping asymmetry $t_2/t_1=0.2$, for $U=0$ and $U=D$. $T^{(1)}$ ($T^{(i)}_{int}$) 
corresponds to the temperature of the non-interacting (interacting) state, with $S/N=1$, while $T^{(f)}$ ($T_{int}^{(f)}$) correspond to that of the final
state (after doing the cooling procedure, and taking $\alpha=0.2$).} 
\end{figure}

Within our approximation, we now attempt to estimate the efficacy of our cooling procedure for the interacting system.
  In Fig. \ref{fig:entropycooling}, we plot the entropy of the interacting ($U$=$D$) and the non-interacting system, for $t_2/t_1$=0.2 at half-filling.  
In this example with $\alpha$=0.2,  we see that starting from $U$=$D$  and entropy per site (at half filling
is identical to per particle) $s$=1, the initial temperature is about 
$T_{int}^{(i)}/D\simeq0.28$. Then, first by decreasing $U$ to zero, we see that it corresponds to a temperature $T^{(1)} \simeq 0.175 D$. 
Following the steps \ref{enum:firststep}-\ref{enum:last}, we can use eq.~\ref{eq:temp} 
to estimate the temperature of the new state. In this example, the 
temperature of the non-interacting system would be $T^{(f)}\simeq 0.11D$, which according to Fig. \ref{fig:entropycooling}, corresponds to an entropy of $s\simeq0.83$. Then, adiabatically turning on the 
interactions to $U$=$D$, it would be possible to get an interacting state that corresponds to a temperature of $T_{int}^{(f)}/D\simeq 0.16$, which is about a 40\% lower than the initial temperature
 However, we reiterate that this number is only an estimate, since we have used Eq.~\ref{eq:temp}
which is accurate only in the linear regime. 
In any case, the percentage could be further increased by increasing the hopping asymmetry, i.e., reducing $\alpha$.
The cooling procedure can be repeated a number of times to access even lower entropies and temperatures. 
It would also be interesting to
see how this scheme  works if we take into account the trapping potential and also consider ordered initial states. This can be done numerically  and is left for future work. 

Regarding the actual implementation of the cooling procedure in experimental setups, we would like to point out that our proposal has  the simple goal of showing that the hopping asymmetry  can potentially be exploited to cool  the system once it has been loaded onto an optical lattice. Though the hopping and interaction parameters can in principle be tuned over a wide range, a main experimental concern is how to remove the excess of entropy from the system in a controlled manner. This is a very challenging issue and is the subject of ongoing work. Another important aspect is the equilibration times of the system. These are topics of great experimental relevance and remain beyond the scope of the present study.

\subsection{Cooling by tuning hopping and interactions}
\label{subsec:coolingbytuning}

The success and applicability of the cooling procedure discussed above depends crucially   on the changes in
temperature induced by the
adiabatic variation of the model parameters in the AHM.  More precisely,  a key issue is whether adiabatic changes
 of  $U$ and  $t_\sigma$ will heat or cool the system\cite{werner_prl95,deleo_pra}.  In this section, we  analyze the temperature changes in the AHM as the parameters $U$ and $r$ are varied.  As in  the previous section, we restrict our study to the disordered phase, but the generalization to a long-range ordered  phase  is straightforward.
From  Refs. \onlinecite{werner_prl95,deleo_pra}, we know that any change of a  parameter $x$  in the Hamiltonian,
will result in   an  entropy change  which is linked to a temperature change as follows:

\begin{equation}
\label{eq:deltas}
-\frac{1}{c}\left.\frac{\delta s}{\delta x}\right|_{T}=\left.\frac{1}{T}\frac{\delta T}{\delta x}\right|_s
\end{equation}
\noindent
where $c=T\delta S/\delta T$ is the specific heat ($c$ is always possitive).

Using  Maxwell's relations, 
\begin{equation}
\left.\frac{\delta s}{\delta U}\right|_{N,t_\sigma,T}=-\left.\frac{\delta d}{\delta T}\right|_{N,t_\sigma,U}
\end{equation}
\begin{equation}
\left.\frac{\delta s}{\delta t_\sigma}\right|_{N,U,t_{-\sigma},T}=-\frac{1}{t_\sigma}\left.\frac{\delta E_{kin,\sigma}}{\delta T}\right|_{N,t_\sigma,U}
\end{equation}
and  \eqref{eq:deltas}, we obtain the following expressions for the change in temperature for fixed entropy, when the Hamiltonian
parameters $U,t_\sigma$ are varied,
\begin{equation}
\label{eq:varu}
\frac{1}{T}\left.\frac{\delta T}{\delta U}\right|_{S,N,t_\sigma}=\frac{1}{c}\left.\frac{\delta d}{\delta T}\right|_{N,t_\sigma,U}
\end{equation}
\begin{equation}
\label{eq:vart}
\frac{1}{T}\left.\frac{\delta T}{\delta t_\sigma}\right|_{S,N,U}=\frac{1}{ct_\sigma}\left.\frac{\delta E_{kin,\sigma}}{\delta T}\right|_{N,t_\sigma,U}
\end{equation}

The changes in temperature when the interaction $U$ is varied is given by the temperature variation of the double occupancy,
whereas, the change in temperature induced by the variation of $t_\sigma$ is dictated by the derivative of the
kinetic energy of the particle species $\sigma$ with respect to the temperature.
As explained in Ref. \onlinecite{deleo_pra}, Eq. \eqref{eq:varu} suggests that increasing the interactions will cool (heat) the system whenever the temperature 
derivative of the double occupancy is negative (positive). In a similar way, Eq. \ref{eq:vart} shows that increasing  the hopping amplitude  will heat (cool) the system if the  temperature derivate of the kinetic energy is positive (negative).

\begin{figure}
\centering
\includegraphics[width=8cm,angle=0]{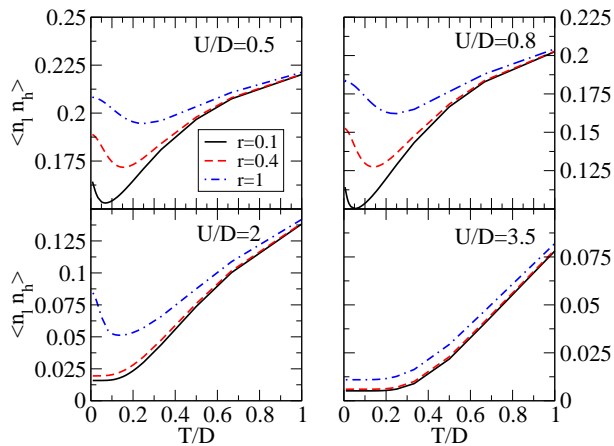}
\caption{\label{fig:doccvst}Double occupancy as a function of $T$, for four values of the interaction (one for each panel), and three different values of the ratio $r$.}
\end{figure}

 We  now analyze  the consequences of tuning theses parameters in the particular case of half-filling without the trapping potential.  
In Fig. \ref{fig:doccvst} we  plot the double occupancy as a function of $T$ for various values of the interaction $U$ and $r$. As  pointed out in Ref. \onlinecite{werner_prl95} for the Hubbard model, at weak coupling, increasing $U$ heats up the system
if one is at high enough $T$ because $\delta d/\delta T >0$.  For $ T < T_m$, where $T_m$ is the temperature at which the double occupancy is the lowest, 
$\delta d/\delta T  < 0$ ,  we have a Pomeranchuk like effect where the system cools with increasing $U$.
This cooling effect is lost at 
 strong couplings,  where the system is insulating and the double occupancy does not vary at low $T$ 
and always has positive slope.  Consequently, there will be neither cooling   nor heating of the system 
whenever $T$ is small, and there will be heating if $T$ is big enough to fill the gap. We find that all of 
these features, first predicted for the HM, 
exist in the AHM case also (cf. top panels of Fig.\ref{fig:doccvst}).  
  
\begin{figure}
\centering
\includegraphics[width=8cm,angle=0]{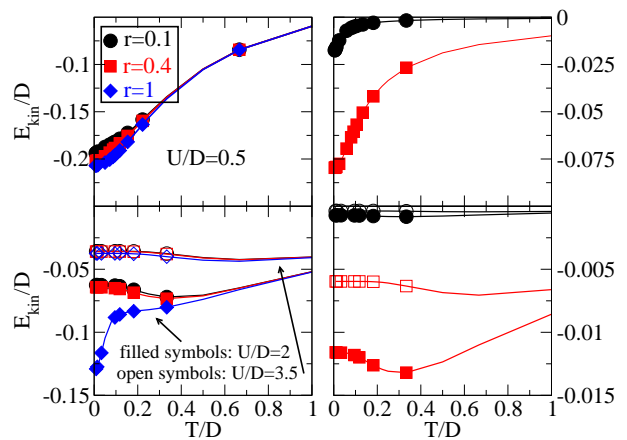}
\caption{\label{fig:ekinvst}Kinetic energy as a function of $T$ for  \up-particles (left panels) and \dw-particles (right panels). Upper panels:  $U/D=0.5$; 
lower panels: $U/D=2$ (filled symbols) and $U/D=3.5$ (empty symbols); black circles, red squares and blue diamonds corresponds to $r=0.1,0.4,1$, respectevely.}
\end{figure}

We now focus on temperature changes induced by the variation of the hoppings. 
In Fig. \ref{fig:ekinvst},  the kinetic energy  for each species is plotted as a function of $T$.  For low $U$, i.e. in the metallic phase, the kinetic energy  of both species increases monotonically with $T$ for all $r$. From Eq.~\ref{eq:vart}, this 
implies that there will always be heating of the system as one increases $t_\sigma$ in the entire temperature range 
of interest. 
In the insulating regime,  since the system is in a Mott state, the kinetic energy remains practically constant at 
low $T$  for both species (especially, the heavier one). Despite the
minor variations seen at higher $T$
  we expect that a change of hopping  parameters at large $U$  leaves the temperature of the system 
effectively unchanged.
  These results  bode well for the cooling procedure outlined  earlier. It would be very interesting to 
use the results of this
  section to calculate the exact change in temperature  resulting from finite variation of the Hamiltonian 
parameters during the
cooling cycle. This is left for future work.

\section{Conclusions}
\label{sec:conclusions}

In the present work we have studied the AHM as a minimal model 
to explore multiorbital correlated physics in cold-atoms. We obtained the phase
diagram including both ordered and disordered phases.  The disordered phase was
found to exhibit an interesting orbital selective crossover at finite temperatures to the non-Fermi liquid
phase  that we first described in Ref. \onlinecite{emilio_prb}.
We showed that the slope of the momentum dependent density distribution 
at the Fermi energy may be a convenient observable to probe and characterize this orbital-selective state at
finite $T$. 
In the ordered phase, which was the main focus of the present study, we showed that
the AHM displays orbital order coexisiting with a charge density wave whenever $r\not=1$. We computed
the behavior of experimentally accesible observables, such as double occupancy, 
which provided clear signatures of the onset of the ordered state.
We also found the regimes of hopping ratio and $U$ where the ordering temperature is highest, that is,
where it may be easier to access the ordered phase.  
We also discussed the relation between  orbital order and the onset of charge density waves, which may  provide
a practical route for manipulating the density distribution inside optical lattices.

Finally, we presented a  new strategy to cool fermions in optical lattices  using   entropic chromatography  
exploiting the features of entropy concentration that follows from hopping asymmetry.  
In fact, currently achievable entropies per particle $s \simeq 1.5$ are still too
high to even observe the intermediate temperature orbital selective state discussed in this paper,  
which is characterized by
an entropy per particle $s \simeq 0.5 - 0.8$. 
This cooling procedure might be a very efficient method to achieve the  low temperatures  required to access exotic
correlated states in  cold fermionic systems in optical lattices. An obvious extension of our work would be to include the trapping potential in the AHM. Though this would  result in quantitative changes to the the ordering temperatures, entropies per particle in
the various phases, etc., we still expect the phase diagram and the physics discussed in this paper to be valid in the
bulk of the system.
We hope that this work may motivate further experimental developments and provide new bridges between
the area of cold atoms and condensed matter physics.

E.A.W. acknowledges support from the French program ANR-09-RPDOC-019-01. 
\appendix
\section{Analytical expressions of the observables}

For completeness, in this appendix we present the different expressions used to compute the physical observables from the DMFT calculation.
The order parameters  are linear combinations of the number of particles, and can be easily computed from the Matsubara Green's functions  obtained by solving the DMFT equations.
 For a given sublattice $\alpha$ and species $\sigma$,
\begin{equation}
\label{eq:nmats}
n_{\alpha\sigma}=\frac{1}{\beta}\sum_{\omega_n=-\infty}^\infty G_{\alpha\sigma}(i\omega_n)\exp(i\omega_n0^+)
\end{equation}
\noindent
Taking into account the problem of convergence involving $|\omega_n|\rightarrow\infty$, Eq. \ref{eq:nmats} can be expressed in the more numerically tractable manner as
\begin{equation}
n_{\alpha\sigma}=\frac{1}{2}+\frac{1}{\beta}\sum_{\omega_n=-\infty}^\infty \textrm{Re}\left[G_{\alpha\sigma}(i\omega_n)\right]
\end{equation}

The thermodynamical observables, however,  require the calculation of the internal energy $E=\langle H\rangle$.
In the disordered phase, the kinetic term is given by
\begin{equation}
E_{\textrm{kin},\sigma}=\frac{1}{\beta}\sum_{k,\omega_n}(\epsilon_{k\sigma}-\mu)G_\sigma(k,i\omega_n)e^{i\omega_n0^+}
\end{equation}
 Within DMFT for the Bethe lattice it can be rewritten as \cite{bible},
\begin{equation}
E_{\textrm{kin},\sigma}=\frac{t_\sigma^2}{\beta}\sum_{\omega_n}\left[G_\sigma(i\omega_n)\right]^2e^{i\omega_n0^+}
\end{equation}
For the ordered phase, this expression can be generalized to
\begin{equation}
E_{\textrm{kin},\sigma}=\frac{t_\sigma^2}{\beta}\sum_{\omega_n}G_{A\sigma}(i\omega_n)G_{B\sigma}(i\omega_n)e^{i\omega_n0^+}
\end{equation}

The potential energy can be simply computed as $U$ times the double occupancy. In a general case away from half-filling, it might be also necessary to add 
a term $E_\mu=\mu(n_h+n_l)$.
The specific heat is then numerically evaluated as
\begin{equation}
C_v=\frac{\partial E}{\partial T}
\end{equation}

The entropy requires the $T$ integration of the specific heat, since at constant volume $TdS=C_vdT$. This can be achieved by integrating $C_v/T$ from $T=0$ or 
$T\rightarrow\infty$. In general, the latter option is better since the asymptotic value $S(T\rightarrow\infty)=N\ln(4)$ is already known, and therefore 
the expression becomes
\begin{equation}
S(T)=S(\infty)-\int_T^\infty dT' \frac{C_v(T')}{T'}
\end{equation}

However, in some cases, such as an ordered state, where there is a singularity in the specific heat, it is also necessary to carry out the integration from $T=0$, 
which  results in 
\begin{equation}
S(T)=S(0)+\int_0^T dT' \frac{C_v(T')}{T'}
\end{equation}
 To use this expression one needs  a good justification for the value of $S(0)$.

\begin{thebibliography}{34}%
\makeatletter
\providecommand \@ifxundefined [1]{%
 \@ifx{#1\undefined}
}%
\providecommand \@ifnum [1]{%
 \ifnum #1\expandafter \@firstoftwo
 \else \expandafter \@secondoftwo
 \fi
}%
\providecommand \@ifx [1]{%
 \ifx #1\expandafter \@firstoftwo
 \else \expandafter \@secondoftwo
 \fi
}%
\providecommand \natexlab [1]{#1}%
\providecommand \enquote  [1]{``#1''}%
\providecommand \bibnamefont  [1]{#1}%
\providecommand \bibfnamefont [1]{#1}%
\providecommand \citenamefont [1]{#1}%
\providecommand \href@noop [0]{\@secondoftwo}%
\providecommand \href [0]{\begingroup \@sanitize@url \@href}%
\providecommand \@href[1]{\@@startlink{#1}\@@href}%
\providecommand \@@href[1]{\endgroup#1\@@endlink}%
\providecommand \@sanitize@url [0]{\catcode `\\12\catcode `\$12\catcode
  `\&12\catcode `\#12\catcode `\^12\catcode `\_12\catcode `\%12\relax}%
\providecommand \@@startlink[1]{}%
\providecommand \@@endlink[0]{}%
\providecommand \url  [0]{\begingroup\@sanitize@url \@url }%
\providecommand \@url [1]{\endgroup\@href {#1}{\urlprefix }}%
\providecommand \urlprefix  [0]{URL }%
\providecommand \Eprint [0]{\href }%
\providecommand \doibase [0]{http://dx.doi.org/}%
\providecommand \selectlanguage [0]{\@gobble}%
\providecommand \bibinfo  [0]{\@secondoftwo}%
\providecommand \bibfield  [0]{\@secondoftwo}%
\providecommand \translation [1]{[#1]}%
\providecommand \BibitemOpen [0]{}%
\providecommand \bibitemStop [0]{}%
\providecommand \bibitemNoStop [0]{.\EOS\space}%
\providecommand \EOS [0]{\spacefactor3000\relax}%
\providecommand \BibitemShut  [1]{\csname bibitem#1\endcsname}%
\let\auto@bib@innerbib\@empty
\bibitem [{\citenamefont {Hubbard}(1963)}]{hubbard-orig}%
  \BibitemOpen
  \bibfield  {author} {\bibinfo {author} {\bibfnamefont {J.}~\bibnamefont
  {Hubbard}},\ }\href {\doibase 10.1098/rspa.1963.0204} {\bibfield  {journal}
  {\bibinfo  {journal} {Proc. Roy. Soc. London}\ }\textbf {\bibinfo {volume}
  {A276}},\ \bibinfo {pages} {238} (\bibinfo {year} {1963})}\BibitemShut
  {NoStop}%
\bibitem [{\citenamefont {Georges}\ \emph {et~al.}(1996)\citenamefont
  {Georges}, \citenamefont {Kotliar}, \citenamefont {Krauth},\ and\
  \citenamefont {Rozenberg}}]{bible}%
  \BibitemOpen
  \bibfield  {author} {\bibinfo {author} {\bibfnamefont {A.}~\bibnamefont
  {Georges}}, \bibinfo {author} {\bibfnamefont {G.}~\bibnamefont {Kotliar}},
  \bibinfo {author} {\bibfnamefont {W.}~\bibnamefont {Krauth}}, \ and\ \bibinfo
  {author} {\bibfnamefont {M.~J.}\ \bibnamefont {Rozenberg}},\ }\href {\doibase
  10.1103/RevModPhys.68.13} {\bibfield  {journal} {\bibinfo  {journal} {Rev.
  Mod. Phys.}\ }\textbf {\bibinfo {volume} {68}},\ \bibinfo {pages} {13}
  (\bibinfo {year} {1996})}\BibitemShut {NoStop}%
\bibitem [{\citenamefont {Falicov}\ and\ \citenamefont
  {Kimball}(1969)}]{fk-orig}%
  \BibitemOpen
  \bibfield  {author} {\bibinfo {author} {\bibfnamefont {L.~M.}\ \bibnamefont
  {Falicov}}\ and\ \bibinfo {author} {\bibfnamefont {J.~C.}\ \bibnamefont
  {Kimball}},\ }\href {\doibase 10.1103/PhysRevLett.22.997} {\bibfield
  {journal} {\bibinfo  {journal} {Phys. Rev. Lett.}\ }\textbf {\bibinfo
  {volume} {22}},\ \bibinfo {pages} {997} (\bibinfo {year} {1969})}\BibitemShut
  {NoStop}%
\bibitem [{\citenamefont {Freericks}\ and\ \citenamefont
  {Zlati\ifmmode~\acute{c}\else \'{c}\fi{}}(2003)}]{fk}%
  \BibitemOpen
  \bibfield  {author} {\bibinfo {author} {\bibfnamefont {J.~K.}\ \bibnamefont
  {Freericks}}\ and\ \bibinfo {author} {\bibfnamefont {V.}~\bibnamefont
  {Zlati\ifmmode~\acute{c}\else \'{c}\fi{}}},\ }\href {\doibase
  10.1103/RevModPhys.75.1333} {\bibfield  {journal} {\bibinfo  {journal} {Rev.
  Mod. Phys.}\ }\textbf {\bibinfo {volume} {75}},\ \bibinfo {pages} {1333}
  (\bibinfo {year} {2003})}\BibitemShut {NoStop}%
\bibitem [{\citenamefont {Si}\ \emph {et~al.}(1992)\citenamefont {Si},
  \citenamefont {Kotliar},\ and\ \citenamefont {Georges}}]{sietal}%
  \BibitemOpen
  \bibfield  {author} {\bibinfo {author} {\bibfnamefont {Q.}~\bibnamefont
  {Si}}, \bibinfo {author} {\bibfnamefont {G.}~\bibnamefont {Kotliar}}, \ and\
  \bibinfo {author} {\bibfnamefont {A.}~\bibnamefont {Georges}},\ }\href
  {\doibase 10.1103/PhysRevB.46.1261} {\bibfield  {journal} {\bibinfo
  {journal} {Phys. Rev. B}\ }\textbf {\bibinfo {volume} {46}},\ \bibinfo
  {pages} {1261} (\bibinfo {year} {1992})}\BibitemShut {NoStop}%
\bibitem [{\citenamefont {Mandel}\ \emph {et~al.}(2003)\citenamefont {Mandel},
  \citenamefont {Greiner}, \citenamefont {Widera}, \citenamefont {Rom},
  \citenamefont {H\"ansch},\ and\ \citenamefont {Bloch}}]{blochprl91}%
  \BibitemOpen
  \bibfield  {author} {\bibinfo {author} {\bibfnamefont {O.}~\bibnamefont
  {Mandel}}, \bibinfo {author} {\bibfnamefont {M.}~\bibnamefont {Greiner}},
  \bibinfo {author} {\bibfnamefont {A.}~\bibnamefont {Widera}}, \bibinfo
  {author} {\bibfnamefont {T.}~\bibnamefont {Rom}}, \bibinfo {author}
  {\bibfnamefont {T.~W.}\ \bibnamefont {H\"ansch}}, \ and\ \bibinfo {author}
  {\bibfnamefont {I.}~\bibnamefont {Bloch}},\ }\href {\doibase
  10.1103/PhysRevLett.91.010407} {\bibfield  {journal} {\bibinfo  {journal}
  {Phys. Rev. Lett.}\ }\textbf {\bibinfo {volume} {91}},\ \bibinfo {pages}
  {010407} (\bibinfo {year} {2003})}\BibitemShut {NoStop}%
\bibitem [{\citenamefont {Taie}\ \emph {et~al.}(2010)\citenamefont {Taie},
  \citenamefont {Takasu}, \citenamefont {Sugawa}, \citenamefont {Yamazaki},
  \citenamefont {Tsujimoto}, \citenamefont {Murakami},\ and\ \citenamefont
  {Takahashi}}]{japanese}%
  \BibitemOpen
  \bibfield  {author} {\bibinfo {author} {\bibfnamefont {S.}~\bibnamefont
  {Taie}}, \bibinfo {author} {\bibfnamefont {Y.}~\bibnamefont {Takasu}},
  \bibinfo {author} {\bibfnamefont {S.}~\bibnamefont {Sugawa}}, \bibinfo
  {author} {\bibfnamefont {R.}~\bibnamefont {Yamazaki}}, \bibinfo {author}
  {\bibfnamefont {T.}~\bibnamefont {Tsujimoto}}, \bibinfo {author}
  {\bibfnamefont {R.}~\bibnamefont {Murakami}}, \ and\ \bibinfo {author}
  {\bibfnamefont {Y.}~\bibnamefont {Takahashi}},\ }\href {\doibase
  10.1103/PhysRevLett.105.190401} {\bibfield  {journal} {\bibinfo  {journal}
  {Phys. Rev. Lett.}\ }\textbf {\bibinfo {volume} {105}},\ \bibinfo {pages}
  {190401} (\bibinfo {year} {2010})}\BibitemShut {NoStop}%
\bibitem [{\citenamefont {Bloch}\ \emph {et~al.}(2008)\citenamefont {Bloch},
  \citenamefont {Dalibard},\ and\ \citenamefont {Zwerger}}]{rmpcoldatoms}%
  \BibitemOpen
  \bibfield  {author} {\bibinfo {author} {\bibfnamefont {I.}~\bibnamefont
  {Bloch}}, \bibinfo {author} {\bibfnamefont {J.}~\bibnamefont {Dalibard}}, \
  and\ \bibinfo {author} {\bibfnamefont {W.}~\bibnamefont {Zwerger}},\ }\href
  {\doibase 10.1103/RevModPhys.80.885} {\bibfield  {journal} {\bibinfo
  {journal} {Rev. Mod. Phys.}\ }\textbf {\bibinfo {volume} {80}},\ \bibinfo
  {pages} {885} (\bibinfo {year} {2008})}\BibitemShut {NoStop}%
\bibitem [{\citenamefont {Cazalilla}\ \emph {et~al.}(2005)\citenamefont
  {Cazalilla}, \citenamefont {Ho},\ and\ \citenamefont
  {Giamarchi}}]{giamarchi}%
  \BibitemOpen
  \bibfield  {author} {\bibinfo {author} {\bibfnamefont {M.~A.}\ \bibnamefont
  {Cazalilla}}, \bibinfo {author} {\bibfnamefont {A.~F.}\ \bibnamefont {Ho}}, \
  and\ \bibinfo {author} {\bibfnamefont {T.}~\bibnamefont {Giamarchi}},\ }\href
  {\doibase 10.1103/PhysRevLett.95.226402} {\bibfield  {journal} {\bibinfo
  {journal} {Phys. Rev. Lett.}\ }\textbf {\bibinfo {volume} {95}},\ \bibinfo
  {pages} {226402} (\bibinfo {year} {2005})}\BibitemShut {NoStop}%
\bibitem [{\citenamefont {Wang}\ \emph {et~al.}(2009)\citenamefont {Wang},
  \citenamefont {Chen},\ and\ \citenamefont {Das~Sarma}}]{sarma}%
  \BibitemOpen
  \bibfield  {author} {\bibinfo {author} {\bibfnamefont {B.}~\bibnamefont
  {Wang}}, \bibinfo {author} {\bibfnamefont {H.-D.}\ \bibnamefont {Chen}}, \
  and\ \bibinfo {author} {\bibfnamefont {S.}~\bibnamefont {Das~Sarma}},\ }\href
  {\doibase 10.1103/PhysRevA.79.051604} {\bibfield  {journal} {\bibinfo
  {journal} {Phys. Rev. A}\ }\textbf {\bibinfo {volume} {79}},\ \bibinfo
  {pages} {051604} (\bibinfo {year} {2009})}\BibitemShut {NoStop}%
\bibitem [{\citenamefont {Dao}\ \emph {et~al.}(2007)\citenamefont {Dao},
  \citenamefont {Georges},\ and\ \citenamefont {Capone}}]{dao}%
  \BibitemOpen
  \bibfield  {author} {\bibinfo {author} {\bibfnamefont {T.-L.}\ \bibnamefont
  {Dao}}, \bibinfo {author} {\bibfnamefont {A.}~\bibnamefont {Georges}}, \ and\
  \bibinfo {author} {\bibfnamefont {M.}~\bibnamefont {Capone}},\ }\href
  {\doibase 10.1103/PhysRevB.76.104517} {\bibfield  {journal} {\bibinfo
  {journal} {Phys. Rev. B}\ }\textbf {\bibinfo {volume} {76}},\ \bibinfo
  {pages} {104517} (\bibinfo {year} {2007})}\BibitemShut {NoStop}%
\bibitem [{\citenamefont {Lin}\ \emph {et~al.}(2006)\citenamefont {Lin},
  \citenamefont {Yi},\ and\ \citenamefont {Duan}}]{lda}%
  \BibitemOpen
  \bibfield  {author} {\bibinfo {author} {\bibfnamefont {G.-D.}\ \bibnamefont
  {Lin}}, \bibinfo {author} {\bibfnamefont {W.}~\bibnamefont {Yi}}, \ and\
  \bibinfo {author} {\bibfnamefont {L.-M.}\ \bibnamefont {Duan}},\ }\href
  {\doibase 10.1103/PhysRevA.74.031604} {\bibfield  {journal} {\bibinfo
  {journal} {Phys. Rev. A}\ }\textbf {\bibinfo {volume} {74}},\ \bibinfo
  {pages} {031604} (\bibinfo {year} {2006})}\BibitemShut {NoStop}%
\bibitem [{\citenamefont {Wille}\ \emph {et~al.}(2008)\citenamefont {Wille},
  \citenamefont {Spiegelhalder}, \citenamefont {Kerner}, \citenamefont {Naik},
  \citenamefont {Trenkwalder}, \citenamefont {Hendl}, \citenamefont {Schreck},
  \citenamefont {Grimm}, \citenamefont {Tiecke}, \citenamefont {Walraven},
  \citenamefont {Kokkelmans}, \citenamefont {Tiesinga},\ and\ \citenamefont
  {Julienne}}]{PhysRevLett.100.053201}%
  \BibitemOpen
  \bibfield  {author} {\bibinfo {author} {\bibfnamefont {E.}~\bibnamefont
  {Wille}}, \bibinfo {author} {\bibfnamefont {F.~M.}\ \bibnamefont
  {Spiegelhalder}}, \bibinfo {author} {\bibfnamefont {G.}~\bibnamefont
  {Kerner}}, \bibinfo {author} {\bibfnamefont {D.}~\bibnamefont {Naik}},
  \bibinfo {author} {\bibfnamefont {A.}~\bibnamefont {Trenkwalder}}, \bibinfo
  {author} {\bibfnamefont {G.}~\bibnamefont {Hendl}}, \bibinfo {author}
  {\bibfnamefont {F.}~\bibnamefont {Schreck}}, \bibinfo {author} {\bibfnamefont
  {R.}~\bibnamefont {Grimm}}, \bibinfo {author} {\bibfnamefont {T.~G.}\
  \bibnamefont {Tiecke}}, \bibinfo {author} {\bibfnamefont {J.~T.~M.}\
  \bibnamefont {Walraven}}, \bibinfo {author} {\bibfnamefont {S.~J. J. M.~F.}\
  \bibnamefont {Kokkelmans}}, \bibinfo {author} {\bibfnamefont
  {E.}~\bibnamefont {Tiesinga}}, \ and\ \bibinfo {author} {\bibfnamefont
  {P.~S.}\ \bibnamefont {Julienne}},\ }\href {\doibase
  10.1103/PhysRevLett.100.053201} {\bibfield  {journal} {\bibinfo  {journal}
  {Phys. Rev. Lett.}\ }\textbf {\bibinfo {volume} {100}},\ \bibinfo {pages}
  {053201} (\bibinfo {year} {2008})}\BibitemShut {NoStop}%
\bibitem [{\citenamefont {Chin}\ \emph {et~al.}(2010)\citenamefont {Chin},
  \citenamefont {Grimm}, \citenamefont {Julienne},\ and\ \citenamefont
  {Tiesinga}}]{RevModPhys.feshbach}%
  \BibitemOpen
  \bibfield  {author} {\bibinfo {author} {\bibfnamefont {C.}~\bibnamefont
  {Chin}}, \bibinfo {author} {\bibfnamefont {R.}~\bibnamefont {Grimm}},
  \bibinfo {author} {\bibfnamefont {P.}~\bibnamefont {Julienne}}, \ and\
  \bibinfo {author} {\bibfnamefont {E.}~\bibnamefont {Tiesinga}},\ }\href
  {\doibase 10.1103/RevModPhys.82.1225} {\bibfield  {journal} {\bibinfo
  {journal} {Rev. Mod. Phys.}\ }\textbf {\bibinfo {volume} {82}},\ \bibinfo
  {pages} {1225} (\bibinfo {year} {2010})}\BibitemShut {NoStop}%
\bibitem [{\citenamefont {Nascimb\`ene}\ \emph {et~al.}(2010)\citenamefont
  {Nascimb\`ene}, \citenamefont {Navon}, \citenamefont {Jiang}, \citenamefont
  {Chevy},\ and\ \citenamefont {Salomon}}]{salomon}%
  \BibitemOpen
  \bibfield  {author} {\bibinfo {author} {\bibfnamefont {S.}~\bibnamefont
  {Nascimb\`ene}}, \bibinfo {author} {\bibfnamefont {N.}~\bibnamefont {Navon}},
  \bibinfo {author} {\bibfnamefont {K.~J.}\ \bibnamefont {Jiang}}, \bibinfo
  {author} {\bibfnamefont {F.}~\bibnamefont {Chevy}}, \ and\ \bibinfo {author}
  {\bibfnamefont {C.}~\bibnamefont {Salomon}},\ }\href {\doibase
  http://dx.doi.org/10.1038/nature08814} {\bibfield  {journal} {\bibinfo
  {journal} {Nature}\ }\textbf {\bibinfo {volume} {463}},\ \bibinfo {pages}
  {1057} (\bibinfo {year} {2010})}\BibitemShut {NoStop}%
\bibitem [{\citenamefont {Dao}\ \emph {et~al.}(2012)\citenamefont {Dao},
  \citenamefont {Ferrero}, \citenamefont {Cornaglia},\ and\ \citenamefont
  {Capone}}]{Capone}%
  \BibitemOpen
  \bibfield  {author} {\bibinfo {author} {\bibfnamefont {T.-L.}\ \bibnamefont
  {Dao}}, \bibinfo {author} {\bibfnamefont {M.}~\bibnamefont {Ferrero}},
  \bibinfo {author} {\bibfnamefont {P.~S.}\ \bibnamefont {Cornaglia}}, \ and\
  \bibinfo {author} {\bibfnamefont {M.}~\bibnamefont {Capone}},\ }\href
  {\doibase 10.1103/PhysRevA.85.013606} {\bibfield  {journal} {\bibinfo
  {journal} {Phys. Rev. A}\ }\textbf {\bibinfo {volume} {85}},\ \bibinfo
  {pages} {013606} (\bibinfo {year} {2012})}\BibitemShut {NoStop}%
\bibitem [{\citenamefont {Alon}\ \emph {et~al.}(2005)\citenamefont {Alon},
  \citenamefont {Streltsov},\ and\ \citenamefont {Cederbaum}}]{ref1}%
  \BibitemOpen
  \bibfield  {author} {\bibinfo {author} {\bibfnamefont {O.~E.}\ \bibnamefont
  {Alon}}, \bibinfo {author} {\bibfnamefont {A.~I.}\ \bibnamefont {Streltsov}},
  \ and\ \bibinfo {author} {\bibfnamefont {L.~S.}\ \bibnamefont {Cederbaum}},\
  }\href {\doibase 10.1103/PhysRevLett.95.030405} {\bibfield  {journal}
  {\bibinfo  {journal} {Phys. Rev. Lett.}\ }\textbf {\bibinfo {volume} {95}},\
  \bibinfo {pages} {030405} (\bibinfo {year} {2005})}\BibitemShut {NoStop}%
\bibitem [{\citenamefont {Scarola}\ and\ \citenamefont
  {Das~Sarma}(2005)}]{ref2}%
  \BibitemOpen
  \bibfield  {author} {\bibinfo {author} {\bibfnamefont {V.~W.}\ \bibnamefont
  {Scarola}}\ and\ \bibinfo {author} {\bibfnamefont {S.}~\bibnamefont
  {Das~Sarma}},\ }\href {\doibase 10.1103/PhysRevLett.95.033003} {\bibfield
  {journal} {\bibinfo  {journal} {Phys. Rev. Lett.}\ }\textbf {\bibinfo
  {volume} {95}},\ \bibinfo {pages} {033003} (\bibinfo {year}
  {2005})}\BibitemShut {NoStop}%
\bibitem [{\citenamefont {Isacsson}\ and\ \citenamefont {Girvin}(2005)}]{ref3}%
  \BibitemOpen
  \bibfield  {author} {\bibinfo {author} {\bibfnamefont {A.}~\bibnamefont
  {Isacsson}}\ and\ \bibinfo {author} {\bibfnamefont {S.~M.}\ \bibnamefont
  {Girvin}},\ }\href {\doibase 10.1103/PhysRevA.72.053604} {\bibfield
  {journal} {\bibinfo  {journal} {Phys. Rev. A}\ }\textbf {\bibinfo {volume}
  {72}},\ \bibinfo {pages} {053604} (\bibinfo {year} {2005})}\BibitemShut
  {NoStop}%
\bibitem [{\citenamefont {Brydon}(2008)}]{brydon}%
  \BibitemOpen
  \bibfield  {author} {\bibinfo {author} {\bibfnamefont {P.~M.~R.}\
  \bibnamefont {Brydon}},\ }\href {\doibase 10.1103/PhysRevB.77.045109}
  {\bibfield  {journal} {\bibinfo  {journal} {Phys. Rev. B}\ }\textbf {\bibinfo
  {volume} {77}},\ \bibinfo {pages} {045109} (\bibinfo {year}
  {2008})}\BibitemShut {NoStop}%
\bibitem [{\citenamefont {Winograd}\ \emph {et~al.}(2011)\citenamefont
  {Winograd}, \citenamefont {Chitra},\ and\ \citenamefont
  {Rozenberg}}]{emilio_prb}%
  \BibitemOpen
  \bibfield  {author} {\bibinfo {author} {\bibfnamefont {E.~A.}\ \bibnamefont
  {Winograd}}, \bibinfo {author} {\bibfnamefont {R.}~\bibnamefont {Chitra}}, \
  and\ \bibinfo {author} {\bibfnamefont {M.~J.}\ \bibnamefont {Rozenberg}},\
  }\href {\doibase 10.1103/PhysRevB.84.233102} {\bibfield  {journal} {\bibinfo
  {journal} {Phys. Rev. B}\ }\textbf {\bibinfo {volume} {84}},\ \bibinfo
  {pages} {233102} (\bibinfo {year} {2011})}\BibitemShut {NoStop}%
\bibitem [{\citenamefont {Camjayi}\ \emph {et~al.}(2007)\citenamefont
  {Camjayi}, \citenamefont {Rozenberg},\ and\ \citenamefont
  {Chitra}}]{camjayi}%
  \BibitemOpen
  \bibfield  {author} {\bibinfo {author} {\bibfnamefont {A.}~\bibnamefont
  {Camjayi}}, \bibinfo {author} {\bibfnamefont {M.~J.}\ \bibnamefont
  {Rozenberg}}, \ and\ \bibinfo {author} {\bibfnamefont {R.}~\bibnamefont
  {Chitra}},\ }\href {\doibase 10.1103/PhysRevB.76.195108} {\bibfield
  {journal} {\bibinfo  {journal} {Phys. Rev. B}\ }\textbf {\bibinfo {volume}
  {76}},\ \bibinfo {pages} {195108} (\bibinfo {year} {2007})}\BibitemShut
  {NoStop}%
\bibitem [{\citenamefont {Gorelik}\ \emph {et~al.}(2010)\citenamefont
  {Gorelik}, \citenamefont {Titvinidze}, \citenamefont {Hofstetter},
  \citenamefont {Snoek},\ and\ \citenamefont {Bl\"umer}}]{prl105_germans}%
  \BibitemOpen
  \bibfield  {author} {\bibinfo {author} {\bibfnamefont {E.~V.}\ \bibnamefont
  {Gorelik}}, \bibinfo {author} {\bibfnamefont {I.}~\bibnamefont {Titvinidze}},
  \bibinfo {author} {\bibfnamefont {W.}~\bibnamefont {Hofstetter}}, \bibinfo
  {author} {\bibfnamefont {M.}~\bibnamefont {Snoek}}, \ and\ \bibinfo {author}
  {\bibfnamefont {N.}~\bibnamefont {Bl\"umer}},\ }\href {\doibase
  10.1103/PhysRevLett.105.065301} {\bibfield  {journal} {\bibinfo  {journal}
  {Phys. Rev. Lett.}\ }\textbf {\bibinfo {volume} {105}},\ \bibinfo {pages}
  {065301} (\bibinfo {year} {2010})}\BibitemShut {NoStop}%
\bibitem [{\citenamefont {Freericks}\ and\ \citenamefont
  {Jarrell}(1995)}]{incomhm}%
  \BibitemOpen
  \bibfield  {author} {\bibinfo {author} {\bibfnamefont {J.~K.}\ \bibnamefont
  {Freericks}}\ and\ \bibinfo {author} {\bibfnamefont {M.}~\bibnamefont
  {Jarrell}},\ }\href {\doibase 10.1103/PhysRevLett.74.186} {\bibfield
  {journal} {\bibinfo  {journal} {Phys. Rev. Lett.}\ }\textbf {\bibinfo
  {volume} {74}},\ \bibinfo {pages} {186} (\bibinfo {year} {1995})}\BibitemShut
  {NoStop}%
\bibitem [{\citenamefont {Freericks}(1993)}]{incomfk}%
  \BibitemOpen
  \bibfield  {author} {\bibinfo {author} {\bibfnamefont {J.~K.}\ \bibnamefont
  {Freericks}},\ }\href {\doibase 10.1103/PhysRevB.47.9263} {\bibfield
  {journal} {\bibinfo  {journal} {Phys. Rev. B}\ }\textbf {\bibinfo {volume}
  {47}},\ \bibinfo {pages} {9263} (\bibinfo {year} {1993})}\BibitemShut
  {NoStop}%
\bibitem [{\citenamefont {van Dongen}(1991)}]{vdongen}%
  \BibitemOpen
  \bibfield  {author} {\bibinfo {author} {\bibfnamefont {P.~G.~J.}\
  \bibnamefont {van Dongen}},\ }\href {\doibase 10.1103/PhysRevLett.67.757}
  {\bibfield  {journal} {\bibinfo  {journal} {Phys. Rev. Lett.}\ }\textbf
  {\bibinfo {volume} {67}},\ \bibinfo {pages} {757} (\bibinfo {year}
  {1991})}\BibitemShut {NoStop}%
\bibitem [{\citenamefont {van Dongen}\ and\ \citenamefont
  {Vollhardt}(1990)}]{vdongenfk}%
  \BibitemOpen
  \bibfield  {author} {\bibinfo {author} {\bibfnamefont {P.~G.~J.}\
  \bibnamefont {van Dongen}}\ and\ \bibinfo {author} {\bibfnamefont
  {D.}~\bibnamefont {Vollhardt}},\ }\href {\doibase
  10.1103/PhysRevLett.65.1663} {\bibfield  {journal} {\bibinfo  {journal}
  {Phys. Rev. Lett.}\ }\textbf {\bibinfo {volume} {65}},\ \bibinfo {pages}
  {1663} (\bibinfo {year} {1990})}\BibitemShut {NoStop}%
\bibitem [{\citenamefont {F\'ath}\ \emph {et~al.}(1995)\citenamefont {F\'ath},
  \citenamefont {Doma\ifmmode~\acute{n}\else \'{n}\fi{}ski},\ and\
  \citenamefont {Lema\ifmmode~\acute{n}\else \'{n}\fi{}ski}}]{fath}%
  \BibitemOpen
  \bibfield  {author} {\bibinfo {author} {\bibfnamefont {G.}~\bibnamefont
  {F\'ath}}, \bibinfo {author} {\bibfnamefont {Z.}~\bibnamefont
  {Doma\ifmmode~\acute{n}\else \'{n}\fi{}ski}}, \ and\ \bibinfo {author}
  {\bibfnamefont {R.}~\bibnamefont {Lema\ifmmode~\acute{n}\else
  \'{n}\fi{}ski}},\ }\href {\doibase 10.1103/PhysRevB.52.13910} {\bibfield
  {journal} {\bibinfo  {journal} {Phys. Rev. B}\ }\textbf {\bibinfo {volume}
  {52}},\ \bibinfo {pages} {13910} (\bibinfo {year} {1995})}\BibitemShut
  {NoStop}%
\bibitem [{\citenamefont {St\"oferle}\ \emph {et~al.}(2006)\citenamefont
  {St\"oferle}, \citenamefont {Moritz}, \citenamefont {G\"unter}, \citenamefont
  {K\"ohl},\ and\ \citenamefont {Esslinger}}]{docc}%
  \BibitemOpen
  \bibfield  {author} {\bibinfo {author} {\bibfnamefont {T.}~\bibnamefont
  {St\"oferle}}, \bibinfo {author} {\bibfnamefont {H.}~\bibnamefont {Moritz}},
  \bibinfo {author} {\bibfnamefont {K.}~\bibnamefont {G\"unter}}, \bibinfo
  {author} {\bibfnamefont {M.}~\bibnamefont {K\"ohl}}, \ and\ \bibinfo {author}
  {\bibfnamefont {T.}~\bibnamefont {Esslinger}},\ }\href {\doibase
  10.1103/PhysRevLett.96.030401} {\bibfield  {journal} {\bibinfo  {journal}
  {Phys. Rev. Lett.}\ }\textbf {\bibinfo {volume} {96}},\ \bibinfo {pages}
  {030401} (\bibinfo {year} {2006})}\BibitemShut {NoStop}%
\bibitem [{\citenamefont {Ho}\ and\ \citenamefont {Zhou}(2009)}]{Ho}%
  \BibitemOpen
  \bibfield  {author} {\bibinfo {author} {\bibfnamefont {T.-L.}\ \bibnamefont
  {Ho}}\ and\ \bibinfo {author} {\bibfnamefont {Q.}~\bibnamefont {Zhou}},\
  }\href {\doibase 10.1073/pnas.0809862105} {\bibfield  {journal} {\bibinfo
  {journal} {Proc. Natl. Acad. Sci. USA}\ }\textbf {\bibinfo {volume} {106}},\
  \bibinfo {pages} {6916} (\bibinfo {year} {2009})}\BibitemShut {NoStop}%
\bibitem [{\citenamefont {Bernier}\ \emph {et~al.}(2009)\citenamefont
  {Bernier}, \citenamefont {Kollath}, \citenamefont {Georges}, \citenamefont
  {De~Leo}, \citenamefont {Gerbier}, \citenamefont {Salomon},\ and\
  \citenamefont {K\"ohl}}]{Bernier}%
  \BibitemOpen
  \bibfield  {author} {\bibinfo {author} {\bibfnamefont {J.-S.}\ \bibnamefont
  {Bernier}}, \bibinfo {author} {\bibfnamefont {C.}~\bibnamefont {Kollath}},
  \bibinfo {author} {\bibfnamefont {A.}~\bibnamefont {Georges}}, \bibinfo
  {author} {\bibfnamefont {L.}~\bibnamefont {De~Leo}}, \bibinfo {author}
  {\bibfnamefont {F.}~\bibnamefont {Gerbier}}, \bibinfo {author} {\bibfnamefont
  {C.}~\bibnamefont {Salomon}}, \ and\ \bibinfo {author} {\bibfnamefont
  {M.}~\bibnamefont {K\"ohl}},\ }\href {\doibase 10.1103/PhysRevA.79.061601}
  {\bibfield  {journal} {\bibinfo  {journal} {Phys. Rev. A}\ }\textbf {\bibinfo
  {volume} {79}},\ \bibinfo {pages} {061601} (\bibinfo {year}
  {2009})}\BibitemShut {NoStop}%
\bibitem [{\citenamefont {Nozi\`eres}(1997)}]{noziere}%
  \BibitemOpen
  \bibfield  {author} {\bibinfo {author} {\bibfnamefont {P.}~\bibnamefont
  {Nozi\`eres}},\ }\href@noop {} {\emph {\bibinfo {title} {Theory of
  {I}nteracting {F}ermi systems}}},\ edited by\ \bibinfo {editor} {\bibnamefont
  {{Westview Press}}}\ (\bibinfo  {publisher} {Westview Press},\ \bibinfo
  {year} {1997})\BibitemShut {NoStop}%
\bibitem [{\citenamefont {Werner}\ \emph {et~al.}(2005)\citenamefont {Werner},
  \citenamefont {Parcollet}, \citenamefont {Georges},\ and\ \citenamefont
  {Hassan}}]{werner_prl95}%
  \BibitemOpen
  \bibfield  {author} {\bibinfo {author} {\bibfnamefont {F.}~\bibnamefont
  {Werner}}, \bibinfo {author} {\bibfnamefont {O.}~\bibnamefont {Parcollet}},
  \bibinfo {author} {\bibfnamefont {A.}~\bibnamefont {Georges}}, \ and\
  \bibinfo {author} {\bibfnamefont {S.~R.}\ \bibnamefont {Hassan}},\ }\href
  {\doibase 10.1103/PhysRevLett.95.056401} {\bibfield  {journal} {\bibinfo
  {journal} {Phys. Rev. Lett.}\ }\textbf {\bibinfo {volume} {95}},\ \bibinfo
  {pages} {056401} (\bibinfo {year} {2005})}\BibitemShut {NoStop}%
\bibitem [{\citenamefont {De~Leo}\ \emph {et~al.}(2011)\citenamefont {De~Leo},
  \citenamefont {Bernier}, \citenamefont {Kollath}, \citenamefont {Georges},\
  and\ \citenamefont {Scarola}}]{deleo_pra}%
  \BibitemOpen
  \bibfield  {author} {\bibinfo {author} {\bibfnamefont {L.}~\bibnamefont
  {De~Leo}}, \bibinfo {author} {\bibfnamefont {J.-S.}\ \bibnamefont {Bernier}},
  \bibinfo {author} {\bibfnamefont {C.}~\bibnamefont {Kollath}}, \bibinfo
  {author} {\bibfnamefont {A.}~\bibnamefont {Georges}}, \ and\ \bibinfo
  {author} {\bibfnamefont {V.~W.}\ \bibnamefont {Scarola}},\ }\href {\doibase
  10.1103/PhysRevA.83.023606} {\bibfield  {journal} {\bibinfo  {journal} {Phys.
  Rev. A}\ }\textbf {\bibinfo {volume} {83}},\ \bibinfo {pages} {023606}
  (\bibinfo {year} {2011})}\BibitemShut {NoStop}%
\end{thebibliography}
%
\end{document}